# Fast and flexible inference for spatial extremes

Peng Zhong[*], Scott A. Sisson and Boris Béranger

May 30, 2025

**Abstract**

Statistical modelling of spatial extreme events has gained increasing attention over the last few decades with max-stable processes, and more recently $r$-Pareto processes, becoming the reference tools for the statistical analysis of asymptotically dependent data. Although inference for $r$-Pareto processes is easier than for max-stable processes, there remain major hurdles for their application to high dimensional datasets within a reasonable timeframe. In addition, both approaches have almost exclusively focused on the Brown-Resnick model, for its Gaussian foundations, and for the continuity of its exponent measure. In this paper, we derive a class of models for which this continuity property holds and present the skewed Brown-Resnick model, an extension of the Brown-Resnick that allows for non-stationarity in the dependence structure, and the truncated extremal-$t$ model, a refinement of the well-known extremal-$t$ model. We use an inference methodology based on the intensity function of the process which is derived from the exponent measure, and demonstrate the statistical and computational efficiency of this approach. Applications to two real-world problems illustrate valuable gains in modelling flexibility as well as appealing computational gains over reference methodologies.

*Keywords:* Spatial extremes, $r$-Pareto process, max-stable process, spectral likelihood.

# 1 Introduction

Statistical modelling of spatial extremes is of major importance for the management and assessment of risks inherent to spatial processes. Modelling the dependence structure in spatial extremes is an active field of research where parametric approaches for asymptotic models face major inferential hurdles and are mostly based on the Brown-Resnick model (Brown and Resnick, 1977; Kabluchko et al., 2009), inhibiting realistic applications in high dimensions, $D$. For example, environmental processes are often observed over large spatial regions and/or with fine resolution, and having $D$ in the order of thousands is not uncommon. In this paper, we demonstrate that these inferential issues can be greatly reduced by using the intensity function as long as the exponent measure is continuous. We establish conditions for this property to hold and leverage it to construct new flexible models.

[*]School of Mathematics and Statistics, UNSW Sydney, Australia.
{Peng.Zhong, Scott.Sisson, B.Beranger}@unsw.edu.au

When extreme events are defined as the maxima recorded over a long period of time, max-stable processes (de Haan, 1984; Schlather, 2002) provide a wide class of models for asymptotically dependent data. However, the likelihood function of such models is intractable in high dimensions unless low-dimensional approximations are used, such as the composite likelihood (Padoan et al., 2010; Davison et al., 2012; Whitaker et al., 2020) and the Vecchia approximation (Huser et al., 2023), which are suitable for dimension up to 1000. A relatively recent alternative spatial modelling framework defines extreme events as the realisations of a process for which a risk functional exceeds some high threshold. The corresponding limiting $r$-Pareto processes (Ferreira and de Haan, 2014; Dombry and Ribatet, 2015; de Fondeville and Davison, 2018; de Fondeville and Davison, 2022) share the same theoretical foundations as max-stable processes. The likelihood for $r$-Pareto processes is more tractable than for max-stable processes but still requires evaluation of an intractable normalising constant, which can be avoided through e.g. score matching (de Fondeville and Davison, 2018; Lederer and Oesting, 2023) or choosing the $L_1$-norm as the risk functional. However, methods based on low-dimensional approximations do not provide reasonable estimates for data with complex dependence structures (nonstationarity, anisotropy) due to their loss of higher-order dependence information, and score matching for $r$-Pareto processes is computationally unstable and might not converge when dependence is weak. Spectral likelihoods are an alternative likelihood-based method that use the likelihood function of a limiting multivariate Poisson point process of the max-stable process (Coles and Tawn, 1991; Engelke et al., 2015; Wadsworth and Tawn, 2014). The same likelihood can also be adopted for the corresponding $r$-Pareto process by choosing the $L_1$-norm as the risk functional. The spectral likelihood is computationally efficient and provides consistent estimates for high-dimensional data with complex dependence structures. However, its computational efficiency needs the assumption that the exponent measure is continuous, which is not always the case for existing models. Likelihood-free alternatives using neural networks (Lenzi et al., 2023) and neural Bayes estimators (Sainsbury-Dale et al., 2024; Richards et al., 2024) scale well with large $D$, but they involve a computationally heavy training phase, require fine model tuning, and do not provide any statistical guarantees. Almost all $r$-Pareto process models focus on the counterpart of the max-stable Brown-Resnick process, which we call the Brown-Resnick $r$-Pareto process, due to its simple intensity function and its greater flexibility over other models (see, e.g., de Fondeville and Davison, 2018; Richards et al., 2024; Hector and Reich, 2024).

Both max-stable and $r$-Pareto processes exhibit asymptotic dependence, meaning they have persisting tail dependence as events become more extreme. Although it can argued that many environmental applications display asymptotic independence (e.g., Bopp et al., 2021; Huser and Wadsworth, 2019) or weakening dependence, it is not always clear whether the process is indeed asymptotically independent or if the spatial dependence has simply weakened to some low but positive level. This justifies the development of more flexible asymptotically dependent models.

The overarching aim of this paper is to define a framework for modelling spatial extremes that allows for a broad class of dependence structures, as well as fast and efficient inference. The contributions of this manuscript are multi-fold: we (a) establish theoretical conditions for max-stable and $r$-Pareto models to have a continuous exponent measure; (b) derive two new max-stable and $r$-Pareto models which offer greater flexibility than Brown-Resnick models in describing anisotropic dependence structures; (c) provide a fast inference methodology using spectral likelihoods that relies on continuity of the exponent measure and is applicable to both model-types; and (d) improve the efficiency of state-of-the-art simulation algorithms for $r$-Pareto processes.

The remainder of this paper is organised as follows: Section 2 provides some background on modelling the dependence structure between spatial extremes using max-stable and $r$-Pareto processes, while Section 3 summarises different strategies to fit parametric models. After these foundations, we provide our theoretical contributions in Section 4, which includes the construction of the skewed Brown-Resnick and truncated extremal-$t$ models. Experiments are conducted on synthetic data in Section 5 to showcase the efficiency of the proposed inference methodology. Finally, two real data analyses involving fitting both max-stable and $r$-Pareto models are presented in Section 6 to illustrate the computational superiority of the proposed methodology over state-of-the-art approaches, and to highlight the gains in flexibility of the dependence structure from the skewed Brown-Resnick model over the Brown-Resnick model.

## 2 Modelling approaches for spatial extremes

A max-stable process, $Z$, with unit Fréchet margins can be characterized as

$$Z(s) = \sup_{i=1}^{\infty} R_i W_i(s), \ s \in \mathcal{S}, \tag{1}$$

where $R_1, R_2, \ldots$, are the points of a Poisson Point Process (PPP) on $(0, \infty)$ and $W_1(s), W_2(s), \ldots$, are independent copies of a stochastic processes $W(s)$ on $\mathcal{S}$ with unit mean (de Haan and Ferreira, 2006; Schlather, 2002). The exponent measure restricted onto $\mathbb{R}_+^D$, denoted by $\kappa$, is given by

$$\kappa\left([0, x]^c\right) = \mathbb{E}\left[\max_{i=1,\ldots,D}\left\{\frac{W(s_i)}{x_i}\right\}\right] = \int_0^\infty 1 - \Pr(W \in [0, xr])\mathrm{d}r, \quad x \in \Omega, \tag{2}$$

where $c$ denotes the complement, $W = (W(s_1), \ldots, W(s_D))^\top$ and $\Omega = \mathbb{R}_+^D \setminus \{0\}$. As such, the distribution function $G$ can be expressed as

$$G(x) = \exp\left\{-\kappa([0, x]^c)\right\} = \exp\left\{-V(x)\right\}, \tag{3}$$

where $V(x)$ is a homogeneous function of order $-1$ called the exponent function. The distribution function $G(x)$ can be also viewed as the probability that no points of the PPP with mean measure $\kappa\left([0, x]^c\right)$ are greater than $x$ at locations $s_i, i = 1, \ldots, D$.

Let $B_D = \{1, \ldots, D\}$ and $B_k = \{b_1, \ldots, b_k\} \subset B_D$, where $b_1 < \cdots < b_k$ such that $x_{B_D} \equiv x$ and $x_{B_k} = (x_{b_1}, \ldots, x_{b_k})$. Let $\Omega_{B_k} = \{x \in \Omega : x_j = 0 \text{ if } j \notin B_k\}$ such that $\partial \Omega = \{\Omega_{B_k}, \forall B_k \text{ and } k = 1, \ldots, D-1\}$ represents the boundaries of $\Omega$, i.e. the set of all subspaces of $\Omega$. The interior of $\Omega$ is denoted by $\Omega^\circ = \{x \in \Omega : x_i > 0, \forall\, i \in B_D\}$, i.e. $\Omega^\circ = \Omega \setminus \partial \Omega$. Depending on the choice of $W$ in (1), the exponent measure $\kappa$ can put mass on both $\partial\Omega$ and $\Omega^\circ$ with the intensity function on each subspace $\Omega_{B_k}$ given by

$$\lim_{x_i \to 0, i \notin B_k} -V_{B_k}(\boldsymbol{x}), \qquad (4)$$

where $V_{B_k} = \frac{\partial^k V}{\partial x_{b_1} \ldots \partial x_{b_k}}$. On $\Omega^\circ$, it can be expressed as $\kappa(\boldsymbol{x}) = -V_{B_D}(\boldsymbol{x})$, where the function $\kappa$ is referred to as the intensity function of the max-stable process.

The theory of $r$-Pareto processes (Dombry and Ribatet, 2015; de Fondeville and Davison, 2022), focuses on the limit distribution of stochastic processes deemed extreme according to some risk functional $r(\cdot)$. Assuming the process $X$ with unit Pareto margins, i.e., $\Pr(X(s) > u) = 1/u, u > 1$, satisfies the regular varying condition (see Dombry and Ribatet, 2015), i.e., $\lim_{u \to \infty} u\Pr(X/u \in B) = \kappa(B)$ for all measurable sets $B \subset C^+(\mathcal{S})$, where $C^+(\mathcal{S})$ is the space of non-negative functions on $\mathcal{S}$, then the limiting process

$$\tilde{Z}(s) = \lim_{u \to \infty} \tfrac{X(s)}{u} | r(\{X(s), s \in \mathcal{S}\}) > u, \qquad (5)$$

defines a simple $r$-Pareto process on $\mathcal{A}_r = \{f \in C^+(\mathcal{S}) : r(f) > 1\}$ with probability measure $\kappa(\cdot \cap \mathcal{A}_r)/\kappa(\mathcal{A}_r)$. Its finite dimensional density is therefore $\frac{\kappa(\boldsymbol{x})}{\kappa(\mathcal{A}_r^D)}$, for $\boldsymbol{x} \in \mathcal{A}_r^D$, where $\kappa$ is the intensity function previously defined and $\mathcal{A}_r^D$ is the set $\mathcal{A}_r$ restricted to $D$ dimensions.

When in the presence of discontinuities in the exponent measure $\kappa$ (i.e., the presence of mass on the boundaries of the set $\mathcal{A}_r^D$, denoted by $\partial\mathcal{A}_r^D$), it means $\tilde{Z}$ can take values on $\partial\Omega \cap \partial\mathcal{A}_r^D$. Inference for $\tilde{Z}$ is therefore challenging since it requires the evaluation of the partial derivatives, $-V_{B_k}(\boldsymbol{x})$. However, the max-stable process $Z$ does not take values on $\partial\Omega$ almost surely since it is defined as pointwise maxima as in (1). If $X$ is in the maximum domain of attraction of $Z$, then as $n \to \infty$, $\{X_i/n, i = 1, \ldots, n\}$ converges to a Poisson process on $\Omega$ with mean measure $\kappa$ (Resnick, 2006, Ch. 6), allowing for modelling the extremes of $X$ via the Poisson points $\{X_i\}$. This approach for modelling $Z$ relies on the convergence of $X$ to the Poisson point process and the convergence of $X$ to the max-stable process by taking pointwise maxima. The fact that $\kappa$ can put mass on $\partial\Omega$ hinders the convergence of $X$, introducing large bias in estimation. In this paper, we provide conditions ensuring the continuity of $\kappa$, and introduce two new models that satisfy them.

Most spatial extremes models are initially constructed as max-stable processes from the spectral representation (1) and their $r$-Pareto equivalent is then derived. For example, letting $W(s) = \exp(Y(s) - \sigma^2/2)$ in (1), where $Y(s)$ is a Gaussian random process with zero mean and variance $\sigma^2 > 0$, yields the popular Brown-Resnick model (Brown and Resnick, 1977; Kabluchko et al., 2009). The corresponding distribution function $G$ is given in Huser and Davison (2013), while Wadsworth and Tawn (2014) provide a closed form expression for $V_{B_k}$ as

$$V_{B_k}(x) = \Phi_{D-k}\left(\log(x_{\overline{B}_k}) - \mu(x_{B_k}); \Gamma\right) \times f(x_{B_k}), \quad x \in \mathbb{R}^D_+,$$

where $\Phi_d(\cdot; \Gamma)$ represents the cumulative distribution function (cdf) of the $d$-dimensional centred Gaussian distribution with positive definite covariance matrix $\Gamma$, $\overline{B}_k = B_D \setminus B_k$, $\mu(x_{B_k})$ is a linear function of $\log(x_{B_k})$, and $f$ is a function of $x_{B_k}$. When $x_{\overline{B}_k} \to 0$, the value of $V_{B_k}$ goes to zero $\forall B_k \neq B_D$, implying the density of the Brown-Resnick model only places mass on $\Omega^\circ$ and consequently that there are no discontinuities in the exponent measure ($\kappa(\partial \mathcal{A}_r) = 0$).

Another construction is to let $W(s) = \tilde{W}(s)^\nu_+/a(s)$, where $\tilde{W}_+(s) = \max(\tilde{W}(s), 0)$, $\tilde{W}(s)$ is a skew-normal process, $a(s) = \mathbb{E}\left[\tilde{W}_+(s)^\nu\right]$ and $\nu > 0$ is the degrees of freedom, yielding the skew extremal-$t$ model (Beranger et al., 2017) which includes the extremal-$t$ model (Opitz, 2013) as a special case. Beranger et al. (2017) derive the analytical expression for the distribution function $G$ and demonstrate the presence of mass in $\partial\Omega$ when $D = 3$. Beranger et al. (2021b) provide the general expression for $V_{B_k}$ (see Thibaud and Opitz, 2015, for results specific to the extremal-$t$), indicating the presence of mass on all subspaces of $\Omega$.

## 3 Inference

The density function $g(\cdot; \theta)$ of most parametric models, with $\theta \in \Theta$, is derived through $D$-fold differentiation of the distribution function (3), resulting in a combinatorial explosion of terms and computational intractability of the likelihood. The most popular solution relies on composite likelihoods (Padoan et al., 2010; Sang and Genton, 2014), with $k$-th order log-likelihood function

$$\ell_{C;k}(\boldsymbol{\theta}; \boldsymbol{x}_1, \ldots, \boldsymbol{x}_n) = \sum_{B_k} \zeta_{B_k} \sum_{i=1}^{n} \log g(\boldsymbol{x}_{i,B_k}; \boldsymbol{\theta}), \quad k < D, \tag{6}$$

where the first summation is over all subsets of $k$ locations indexed by $B_k$, $\boldsymbol{x}_{i,B_k}$ extracts the $B_k$ components of $\boldsymbol{x}_i$, and $\zeta_{B_k} \geq 0$ are weights often set as binary values based on a fixed cut-off distance between sites. Practical implementations have mostly focused on pairs or triples (Padoan et al., 2010; Huser and Davison, 2013) while Castruccio et al. (2016) considered higher orders up to $k = 11$. The above works conclude that the pairwise likelihood (i.e, $k = 2$) offers the best compromise between statistical and computational efficiency. Alternatives include the Stephenson-Tawn likelihood, which makes use of the knowledge of occurrence times of maxima (Stephenson and Tawn, 2005; Wadsworth and Tawn, 2014), and which can be combined with the composite likelihood (Beranger et al., 2021b). The methodology of Wadsworth (2015) reduces the potentially large bias of the resulting estimators, for moderate dimension $D$. Overall, these approaches require additional information, are limited to moderate dimensions, and/or exhibit bias.

Inference for max-stable processes can also be performed via the spectral likelihood introduced by Coles and Tawn (1991) where the data $\{\boldsymbol{X}_i, i = 1, \ldots, n\}$ that are in the domain of attraction of $\boldsymbol{Z}$ can be approximately treated as points of a PPP with measure $\kappa(\cdot)$. Assuming the approximation valid for $A_u = \{\boldsymbol{x} \in \mathbb{R}_+^D : \|\boldsymbol{x}\|_1 > u\}$, $u > 0$, the spectral likelihood is given by

$$L(\boldsymbol{x}_1, \ldots, \boldsymbol{x}_{n_A}) = \exp\{-\kappa_D(A)\} \prod_{i=1}^{n_A} \kappa(\boldsymbol{x}_i), \tag{7}$$

where $n_A$ is the number of points in $A_u$. For parametric models, the spectral log-likelihood is

$$\ell_A(\boldsymbol{\theta}; \boldsymbol{x}_1, \ldots, \boldsymbol{x}_n) \propto \sum_{i \in \{m : \|\boldsymbol{x}_m\|_1 > u\}} \log \kappa(\boldsymbol{x}_i; \boldsymbol{\theta}). \tag{8}$$

for some high enough threshold $u$. This approach has been applied to several multivariate extreme value models (see, e.g., Cooley et al., 2010; Engelke et al., 2015; Beranger and Padoan, 2015; Sabourin et al., 2013) and in the spatial context to the Brown-Resnick model (Engelke et al., 2015).

Unlike composite likelihood estimates which are asymptotically unbiased and normally distributed, spectral likelihood estimates are biased since the likelihood function is not the likelihood of the max-stable process itself but of a limiting Poisson process. Bias occurs when events with very small components fall into the region $A_u$, which can be reduced by introducing censoring (see Wadsworth and Tawn, 2014; Thibaud and Opitz, 2015), and when the approximation is poor due to non-zero mass for $\kappa$ on the boundaries. In Section 5.2, simulation experiments for models with continuous $\kappa$ indicate that the bias can be controlled to an acceptable level by appropriately selecting a threshold $u$, and is balanced out by significant computational gains. The likelihood (8) requires the exponent measure to place all its mass in the interior of its domain (Thibaud and Opitz, 2015), like the Brown-Resnick model but not the skew extremal-$t$. We establish general conditions ensuring this property holds in Section 4, and construct new models that satisfy them.

For $r$-Pareto processes with extremeness defined through $r(x_i) > u, i = 1, \ldots, n$, the log-likelihood is written as

$$\ell_{rP}(\theta; x_1, \ldots, x_n) = \sum_{i \in \{m: r(x_m) > u\}} \log\left(\frac{\kappa(z_i; \theta)}{\kappa(\mathcal{A}_r; \theta)}\right), \tag{9}$$

where we recall that $\mathcal{A}_r = \{f \in C_+(\mathcal{S}) : r(f) \geq 1\}$, and $z_i = x_i/u$ represent the realizations of the $r$-Pareto process. The normalising constant, $\kappa(\mathcal{A}_r; \theta)$, involves the evaluation of integrals in $\mathbb{R}_+^D$, leading to intractability of the likelihood as the dimension $D$ increases. Simplifications occur when $r(x) = x(s)$ or $r(x) = \sum_{j=1}^D x(s_j)/D$ since the constant does not then depend on the model parameters, or when $r(x) = \max(x)$ (see de Fondeville and Davison, 2018; Dombry et al., 2024). The spectral likelihood (7) can be seen as a special case of a $r$-Pareto model with risk function $r(x) = \|x\|_1$. To circumvent the difficulties of computing the likelihood of a $r$-Pareto model, de Fondeville and Davison (2018) propose a score matching approach that avoids evaluation of $\kappa(\mathcal{A}_r; \theta)$. Although this approach has mainly been applied to the Brown-Resnick model, it can be potentially used for any model with discontinuities in the exponent measure (such as the skew extremal-$t$), given a careful selection of a weight function to downweight the partial derivative terms in (4) evaluated on the boundaries.

As suggested by Dombry et al. (2024, Proposition 2) and Theorem 4, using rejection sampling, one can generate samples from a $r$-Pareto process with risk functional $r_2$ from samples of a $r$-Pareto process associated with risk functional $r_1$ as long as $Mr_1(\cdot) \geq r_2(\cdot)$, $M > 0$. Following from this fact, and that the selected data in (9) follow the $r$-Pareto distribution with the $L_1$ norm risk functional ($\equiv r_0(\cdot)$), we propose to use the likelihood of the $r_0$-Pareto process to make inference about any $r$-Pareto process with a different risk functional by choosing a high threshold $u > M$ in (9). This particularly applies to $L_p$ norms, $p > 1$, since $L_p$ bounds $L_1$ for finite $p$. Indeed, from Hölder's inequality for $L_p$ norms we have $r_0(\cdot) = \|\cdot\|_1 \leq D^{1-1/p}\|\cdot\|_p$, $p > 1$, and therefore can choose $u > D^{1-1/p}$, $p > 1$, to infer the $r$-Pareto process with the $L_p$ norm risk functional using (8) and produce unbiased parameter estimates. This approach to fitting $r$-Pareto processes with different risk functionals than the $L_1$ norm considerably reduces the computational complexity of evaluating $\ell_{rP}$ since it avoids computation of the normalising constant and, to our knowledge, has not previously been considered in the literature.

As previously mentioned, in the case of model misspecification and where the data only falls into the max-domain of attraction of the model, a censored likelihood approach can help mitigate bias (Wadsworth and Tawn, 2014; Huser et al., 2016) but is limited to $D \leq 30$. When the censored likelihood is not practical and model misspecification is a significant concern, Huser et al. (2016) suggest that a block maxima approach is preferred since it will artificially increase the tail dependence strength between locations and push the data further towards the fitted dependence model. As such, the simulation experiments of Section 5 focus on comparing the performance of the composite and the spectral likelihood for max-stable processes, while for $r$-Pareto processes, the performance of the spectral likelihood will be compared against the score matching method.

# 4 Theoretical results

## 4.1 Restricting spatial extreme models to $\Omega^\circ$

To capitalise on the computational potential of the spectral likelihood (7), we establish general conditions for any max-stable and $r$-Pareto model to concentrate their mass on $\Omega^\circ$. As $r$-Pareto models are most often derived from their max-stable counterpart (Section 2), our approach follows a max-stable model perspective.

From the exponent function (2), the max-stable process intensity function can be expressed as

$$\kappa(\boldsymbol{x}) = -V_{B_D}(\boldsymbol{x}) = \int_0^\infty r^D f_{\boldsymbol{W}}(\boldsymbol{x}r)\mathrm{d}r, \qquad (10)$$

where $f_{\boldsymbol{W}}$ is the joint density function of $\boldsymbol{W}$ and, without loss of generality, the partial derivatives of $V(\boldsymbol{x})$ with respect to the components in $B_k$ can be obtained through

$$-V_{B_k}(\boldsymbol{x}) = \int_0^\infty \frac{\partial^k}{\partial x_1 \partial x_2 \ldots \partial x_k} \Pr(\boldsymbol{W} \in [\boldsymbol{0}, \boldsymbol{x}r])\mathrm{d}r. \qquad (11)$$

By decomposing the joint probability function $\Pr(W \in [\boldsymbol{0}, \boldsymbol{x}r])$ into a conditional and a marginal component, and assuming its partial derivatives do exist, (11) can be re-written as

$$-V_{B_k}(\boldsymbol{x}) = \int_0^\infty r^k f_{\boldsymbol{W}_{B_k}}(r\boldsymbol{x}_{B_k})\Pr(\boldsymbol{W}_{\overline{B}_k} \in [\boldsymbol{0}_{D-k}, r\boldsymbol{x}_{\overline{B}_k}] \mid \boldsymbol{W}_{B_k} = r\boldsymbol{x}_{B_k})\mathrm{d}r, \qquad (12)$$

where $\boldsymbol{0}_{D-k}$ is the $D-k$ sub-vector of the zero vector $\boldsymbol{0}$. From (12), note that if the conditional probability $\Pr(\boldsymbol{W}_{\overline{B}_k} = \boldsymbol{0}_{D-k} \mid \boldsymbol{W}_{B_k} = \boldsymbol{x}_{B_k}) = 0$ for all $k, 1 \leq k \leq D-1$ and any $\boldsymbol{x}_{B_k} > \boldsymbol{0}_k$, then the intensity function in (4) has no mass on $\partial\Omega$. The following theorem formalises this condition.

**Theorem 1.** *Let a max-stable process $\{Z(\boldsymbol{s}), \boldsymbol{s} \in \mathcal{S}\}$ be defined as in (1) with a smooth exponent function $V$ at any finite set of $D$ locations such that the partial derivatives of the function $V$ exist. The intensity function on $\partial\Omega$, as defined in (4), is zero almost everywhere if and only if the conditional probability of $\boldsymbol{W}$ satisfies*

$$\Pr(\boldsymbol{W}_{\overline{B}_k} = \boldsymbol{0}_{D-k} \mid \boldsymbol{W}_{B_k} = \boldsymbol{x}_{B_k}) = 0, \ \forall \ k \in \{1, \ldots, D-1\}, \ \boldsymbol{x}_{B_k} > \boldsymbol{0}_k \ \textit{almost everywhere}. \qquad (13)$$

The proof can be found in Section A of the Supplemental Material. This result indicates that the behavior of $V_{B_k}$ on $\partial\Omega$ is determined by the behavior of the process $W$ on its lower-end boundary at zero. Section 2 states that for the Brown-Resnick model, $W = \exp(\tilde{W} - \sigma^2/2)$ with $\tilde{W}$ a centered Gaussian random process. The event $\left\{W_{\overline{B}_k} = \mathbf{0}_{D-k} \mid W_{B_k} = \boldsymbol{x}_{B_k}\right\}$ is equivalent to the event $\left\{\tilde{W}_{\overline{B}_k} = -\infty_{D-k} \mid \tilde{W}_{B_k} = \log(\boldsymbol{x}_{B_k}) + \sigma^2/2\right\}$, and since the conditional distribution of $\tilde{W}_{\overline{B}_k} \mid \tilde{W}_{B_k}$ is Gaussian, then (13) holds, Theorem 1 confirms that the Brown-Resnick model does not put mass on $\partial\Omega$. Section 2 also states that for the skewed extremal-$t$ model, we have $W = \max(\tilde{W}^\nu, 0)$ with $\tilde{W}$ a skew-normal process and $\nu > 0$. The event $\{W_{\overline{B}_k} = \mathbf{0}_{D-k} \mid W_{B_k} = r\boldsymbol{x}_{B_k}\}$ is equivalent to the event $\left\{\tilde{W}_{\overline{B}_k} \leq \mathbf{0}_{D-k} \mid \tilde{W}_{B_k} = (r\boldsymbol{x}_{B_k})^{1/\nu}\right\}$, which is not a null event and so the skewed extremal-$t$ model does indeed put mass on $\partial\Omega$.

## 4.2 Extending current classes of max-stable and $r$-Pareto models

We now take advantage of Theorem 1 to construct spatial models that only put mass on $\Omega^\circ$, and which can thereby produce tractable likelihood functions.

### 4.2.1 The skewed Brown-Resnick process

Following the Brown-Resnick construction, to develop max-stable processes with no mass on $\partial\Omega$, one can take the exponential marginal transformation of a process with a light marginal tail that places zero mass on the infimum of its marginal space. The following theorem extends the Brown-Resnick process by using the skew-normal distribution for $Y(\boldsymbol{s})$.

**Theorem 2.** *Let $W(\boldsymbol{s}) = \exp\{Y(\boldsymbol{s}) - a(\boldsymbol{s})\}$ where $Y(\boldsymbol{s})$ is a centred skew-normal process with scale matrix $\Sigma$ and slant parameter $\boldsymbol{\alpha}$, and $a(\boldsymbol{s}) = \log\mathbb{E}\left[\exp\{Y(\boldsymbol{s})\}\right]$. The corresponding max-stable process defined via (1) is called the skewed Brown-Resnick process and is characterized by the following exponent function:*

$$V(\boldsymbol{x}) = \sum_{k=1}^{D} \frac{1}{x_k} \Psi\left(\log\left(\frac{\boldsymbol{x}^{\circ\circ}_{-k}}{x^{\circ\circ}_k}\right) + \frac{\tilde{\omega}^2_{-k} - \tilde{\omega}^2_k \mathbf{1}_{D-1}}{2}; \boldsymbol{\mu}_k, \Sigma_k, \widehat{\boldsymbol{\alpha}}_k, \tau_k\right), \tag{14}$$

*where $\Psi$ denotes the cdf of the extended skew-normal distribution (see Definition 1 in Section B of the Supplemental Material), $x^{\circ\circ} = (x_1^{\circ\circ}, \ldots, x_D^{\circ\circ})$, $x_k^{\circ\circ} = 2x_k \Phi(\tau_k; 0, 1)$, $k = 1, \ldots, D$, $\tau_k = \tilde{\omega}_k \delta_k$, $\tilde{\omega} = (\tilde{\omega}_1, \ldots, \tilde{\omega}_D) = \omega \mathbf{1}$, $\omega = \sqrt{\mathrm{diag}(\Sigma)}$, $\delta = \left(1 + \alpha^\top \overline{\Sigma} \alpha\right)^{-1/2} \overline{\Sigma} \alpha$, $\overline{\Sigma} = \omega^{-1} \Sigma \omega$, $\mu_k = A_k \Sigma_{\cdot k}$, $A_k = (e_1, \ldots, e_{k-1}, -\mathbf{1}, e_k, \ldots, e_{D-1})$, $e_k$, $k = 1, \ldots, D-1$ are standard basis vectors of dimension $(D-1)$, $\Sigma_k = A_k \Sigma A_k^\top$, $\widehat{\alpha}_k = \left(1 - \delta^\top \omega A_k^\top \Sigma_k^{-1} A_k \omega \delta\right)^{-1/2} \widehat{\omega}_k \Sigma_k^{-1} A_k \omega \delta$, and $\widehat{\omega}_k = \sqrt{\mathrm{diag}(\Sigma_k)}$.*

**Proposition 1.** *The skewed Brown-Resnick model defined in Theorem 2 has zero mass on $\partial \Omega$ and its density on $\Omega^\circ$ is given by*

$$\kappa(x) = \frac{2\Phi(\tilde{\tau}; 0, 1) |\Sigma|^{-1/2} (\mathbf{1}^\top q)^{-1/2}}{(2\pi)^{(D-1)/2} \prod_{k=1}^D w_k} \exp\left\{-\frac{1}{2}\left[\log x^{\circ\circ \top} \mathcal{M} \log x^{\circ\circ} + \log x^{\circ\circ \top} \left(\frac{2q}{\mathbf{1}^\top q} + \mathcal{M}\tilde{\omega}^2\right) + \frac{q^\top \tilde{\omega}^2 - 1}{\mathbf{1}^\top q} + \frac{1}{4}\tilde{\omega}^{2,\top} \mathcal{M} \tilde{\omega}^2\right]\right\}, \quad (15)$$

*where $q = \Sigma^{-1} \mathbf{1}$, $\mathcal{M} = \Sigma^{-1} - qq^\top / \mathbf{1}^\top q$ and*

$$\tilde{\tau} = \left(1 + \frac{(\alpha^\top \omega^{-1} \mathbf{1})^2}{\mathbf{1}^\top q}\right)^{-1/2} \alpha^\top \omega^{-1} \left[\left(\mathbb{I} - \frac{\mathbf{1} q^\top}{\mathbf{1}^\top q}\right) \left(\log x^{\circ\circ} + \frac{\tilde{\omega}^2}{2}\right) + \frac{\mathbf{1}}{\mathbf{1}^\top q}\right].$$

Setting $\alpha = 0$ implies $\widehat{\alpha}_k = 0$, $\tau_k = 0$ and therefore $x^{\circ\circ} = x$, recovering the Brown-Resnick model. Furthermore, as argued by Azzalini (2013, Section 5.2), inference involving the extended skew-normal distribution can be improved through the re-parametrisation $\eta = \omega^{-1} \alpha$. As such, we have $\omega \delta = \Sigma \eta (1 + \eta^\top \Sigma \eta)^{-1/2} \equiv \xi$, and in (14) and (15) we get $\tau_k = \xi_k$ and $\widehat{\alpha}_k = \left(1 - \xi^\top A_k^\top \Sigma_k^{-1} A_k \xi\right)^{-1/2} \widehat{\omega}_k \Sigma_k^{-1} A_k \xi$.

Note that both matrices $\mathcal{M}$ and $\mathbb{I} - \frac{\mathbf{1} q^\top}{\mathbf{1}^\top q}$ given in Proposition 1 are singular with a null space $\{c\mathbf{1} : c \in \mathbb{R}\}$, which makes the model less identifiable for the slant parameter $\alpha$, in particular when using the intensity function. We therefore impose the condition $\eta^\top \mathbf{1} = \alpha^\top \omega^{-1} \mathbf{1} = 0$ to improve the identifiability of the skewed Brown-Resnick process. Similar conditions were applied by Padoan (2011) and Beranger et al. (2019) when considering skewed versions of the Hüsler-Reiss model, the multivariate equivalent of the Brown-Resnick model.

The proofs of Theorem 2 and Proposition 1 are relegated to Section C.1 and C.2 of the Supplemental material, where the above re-parametrisation is also introduced to highlight the reduction in computational complexity of evaluating (14). Section C.3 provides the partial derivatives of the exponent function, required to compute the density of the corresponding $r$-Pareto process.

### 4.2.2 The truncated extremal-$t$ process

This section introduces a modification of the extremal-$t$ to remove its density mass on $\partial\Omega$. Similar results can be derived for the skewed extremal-$t$.

**Theorem 3.** *Let $W(s) = \tilde{Y}(s)^\nu/a(s)$, where $\tilde{Y}(s) = Y(s)|Y(s) > 0$, $Y(s)$ is a centred Gaussian process with unit variances and correlation matrix $\Sigma$, and $a(s) = \mathbb{E}\left[\tilde{Y}(s)^\nu\right]$, $\nu > 0$. The corresponding max-stable process defined via (1) is called the truncated extremal-t process and is characterized by the following exponent function:*

$$V(\boldsymbol{x}) = \sum_{k=1}^{D} \frac{T^\circ_{\nu+1}\left(\left(\frac{\boldsymbol{x}^\circ_{-k}}{\boldsymbol{x}^\circ_k}\right)^{1/\nu};\Sigma_{-k,k},\frac{\mathbb{I}_{D-1}-\Sigma_{-k,k}\Sigma_{k,-k}}{\nu+1}\right)}{\boldsymbol{x}_k T^\circ_{\nu+1}\left(\infty_{D-1};\Sigma_{-k,k},\frac{\mathbb{I}_{D-1}-\Sigma_{-k,k}\Sigma_{k,-k}}{\nu+1}\right)}, \qquad (16)$$

*where $T^\circ_\nu(\cdot;\mu,\Sigma)/T^\circ_\nu(\infty;\mu,\Sigma)$ denotes the cdf of the multivariate truncated student-$t$ distribution with mean $\mu \in \mathbb{R}^D$, positive scale matrix $\Sigma$ and degrees of freedom $\nu$ (see Definition 4 in Section B of the Supplemental Material), and $\boldsymbol{x}^\circ = \boldsymbol{x}\boldsymbol{a}$ with*

$$a_k = \frac{2^{(\nu-2)/2}\Gamma((\nu+1)/2)}{\Phi^\circ(\infty;0,\Sigma)\sqrt{\pi}} T^\circ_{\nu+1}\left(\infty_{D-1};\Sigma_{-k,k},\frac{\mathbb{I}_{D-1}-\Sigma_{-k,k}\Sigma_{k,-k}}{\nu+1}\right), \quad k = 1,\ldots,D,$$

*$\mathbb{I}_{D-1}$ is the $(D-1)$-dimensional identity matrix, and the index $-k$ represents the set $B_D\setminus\{k\}$.*

Note that the first term in $a_k$ is a constant and can be omitted when evaluating (16).

**Proposition 2.** *The truncated extremal-t model defined in Theorem 3 has zero mass on $\partial\Omega$ and its density on $\Omega^\circ$ is given by*

$$\kappa(\boldsymbol{x}) = \frac{\prod_{k=1}^{D}\left(a_k w_k^{1-\nu}\right)^{1/\nu}}{\Phi^\circ(\infty;0,\Sigma)\nu^{D-1}|\Sigma|^{1/2}\pi^{D/2}2^{(2-\nu)/2}}\left\{\left[(\boldsymbol{x}^\circ)^{1/\nu}\right]^\top \Sigma^{-1}\left[(\boldsymbol{x}^\circ)^{1/\nu}\right]\right\}^{-(\nu+D)/2}\Gamma\left(\frac{\nu+D}{2}\right). \qquad (17)$$

Proofs of Theorem 3 and Proposition 2 are given in Section D.1 and D.2 of the Supplemental Material, while the partial derivatives of the exponent function, required to compute the density of the corresponding $r$-Pareto process, are given in Section D.3. The main computational bottleneck in computing (17) resides in evaluating the normalizing constants $a$ which involves the multivariate truncated Gaussian and Student-$t$ distribution functions, but solutions of up to $D = 1000$ dimensions are offered by the R package TruncatedNormal (Botev and Belzile, 2021).

## 4.3 Dependence structure flexibility

To measure the strength of the extremal dependence of a max-stable model, one choice relies on the probability that all $D$ components of $Z$ are less than some value $z > 0$, written as

$$G(z) = \Pr(Z(s_1) < z, \ldots, Z(s_D) < z) = \exp\{-z^{-1}V(1)\} = \exp\{-z^{-1}\theta_D\}, z = (z, \ldots, z),$$

where $\theta_D \in [1, D]$ is called the extremal coefficient, with the lower and upper bounds respectively indicating full dependence and independence (see Smith, 1990; Davison et al., 2012). For spatial locations $(s_1, s_2)$, the extremal coefficient can be formulated using the limiting conditional probability of exceedance, i.e., $\theta_2 = 2 - \lim_{z \to \infty} \Pr(Z(s_1) > z | Z(s_2) > z)$. For $r$-Pareto processes, de Fondeville and Davison (2018) define the bivariate extremal coefficient of $\tilde{Z}$, defined in (5), as

$$\tilde{\theta}_2 = \Pr\left(\tilde{Z}(s_1) > u | \tilde{Z}(s_2) > u, r(\tilde{Z}) > 1\right), \quad \forall (i, j) \in B_D,$$

and when $\{\tilde{Z} : r(\tilde{Z}) \geq 1\} \subseteq \{\tilde{Z} : Z(s_1) > u, Z(s_2) > u\}$ (occurring as $u \to \infty$), then $\tilde{\theta}_2 = 2 - \theta_2$.

For a skewed Brown-Resnick model observed at two locations ($s_1$ and $s_2$), with covariance matrix $\Sigma = [\sigma_{ij}]_{i,j=1,2}$, the extremal coefficient is

$$\theta_2^{\text{sBR}} = \sum_{i,j=1,2, i \neq j} \Phi^{-1}(\xi_i) \Phi\left(\begin{bmatrix} \log\left(\frac{\Phi(\xi_j;0,1)}{\Phi(\xi_i;0,1)}\right) + \gamma_{12} \\ \xi_i \end{bmatrix}; \begin{bmatrix} 0 \\ 0 \end{bmatrix}, \begin{bmatrix} 2\gamma_{12} & \xi_i - \xi_j \\ \xi_i - \xi_j & 1, \end{bmatrix}\right), \quad (18)$$

where $\gamma_{12} = (\sigma_{11} + \sigma_{22} - 2\sigma_{12})/2$. In addition to $\Sigma$, (18) is also controlled by $\xi$ which has a one-to-one relationship with $\eta$ via $\eta = (1 - \xi \Sigma^{-1} \xi)^{-1/2} \Sigma^{-1} \xi$. When $\xi = 0$, (18) becomes $\theta_2^{\text{BR}} = 2\Phi(\sqrt{\gamma_{12}/2}; 0, 1)$, the extremal coefficient of the Brown-Resnick model. Similarly, the bivariate extremal coefficient for the truncated extremal-$t$ with correlation $\rho \in [-1, 1]$, is

$$\theta_2^{\text{tET}} = 2 \frac{T_{\nu+1}\left((1-\rho)\sqrt{\frac{\nu+1}{1-\rho^2}}\right) - T_{\nu+1}\left(-\rho\sqrt{\frac{\nu+1}{1-\rho^2}}\right)}{1 - T_{\nu+1}\left(-\rho\sqrt{\frac{\nu+1}{1-\rho^2}}\right)}, \quad (19)$$

which can also be deduced from Beranger et al. (2021a) and from which we recognize the extremal coefficient of the extremal-$t$ model $\theta_2^{\text{ET}} = 2T_{\nu+1}\left((1-\rho)\sqrt{\frac{\nu+1}{1-\rho^2}}\right)$. Figure 1a illustrates the range of extremal dependence structures obtained from the skewed Brown-Resnick model, by displaying $\theta_2^{\text{sBR}}$, with $\boldsymbol{\eta} = (\eta_1, -\eta_1)^\top$ and $\Sigma = ((\sigma_{22}, \sigma_{12})^\top, (\sigma_{12}, \sigma_{22})^\top)$. For fixed $\Sigma$, the $\eta$ parameter, which essentially controls the skewness, can increase the level of dependence compared to the Brown-Resnick case ($\eta = 0$). A similar observation is made from Figure 1b where the truncated extremal-$t$ process (blue) exhibits stronger dependence between components than the extremal-$t$ process (red), although the gap decreases with increasing degree of freedom, $\nu$. This increase in the dependence strength is a direct consequence of the removal of the mass on $\partial\Omega$. In addition, Figure 1a also showcases flexibility gains with increasing $\sigma_{22}$ (left to right panel).

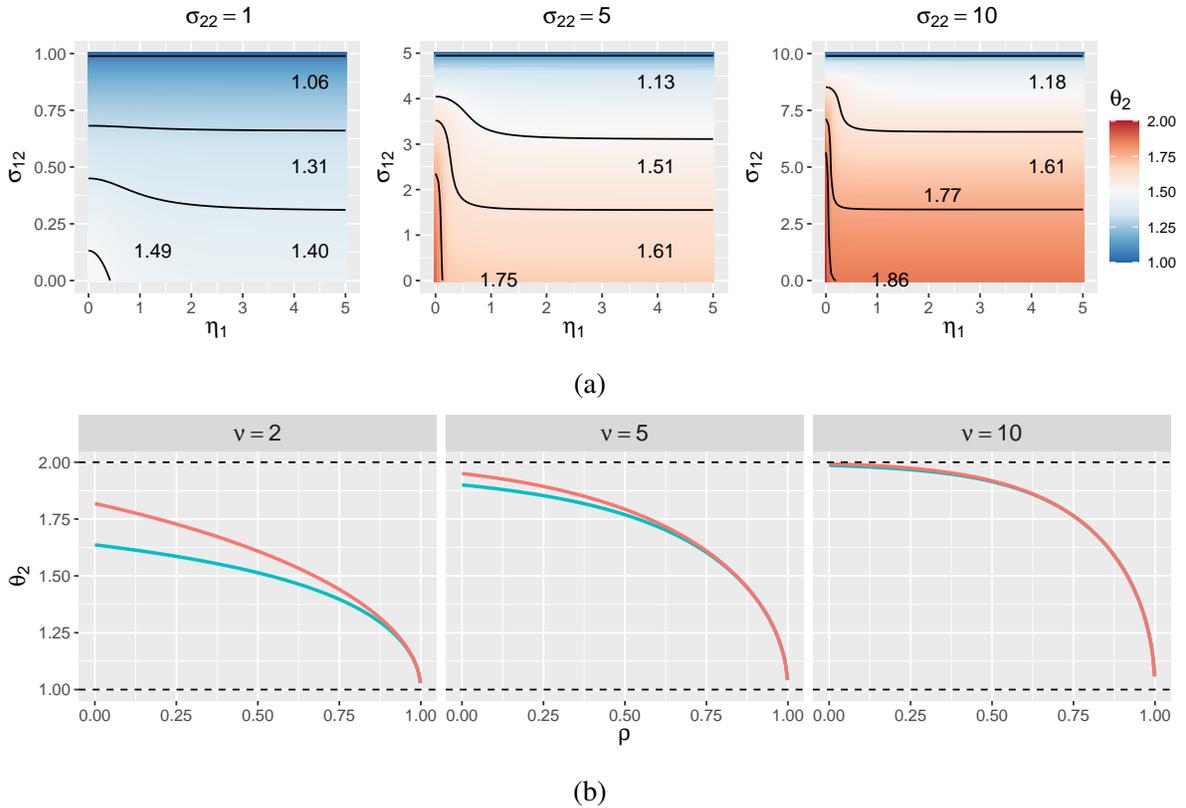

Figure 1: (a): Bivariate extremal coefficient of the skewed Brown-Resnick model ($\theta_2^{\text{sBR}}$) with $\boldsymbol{\eta} = (\eta_1, -\eta_1)^\top$ and $\Sigma = ((\sigma_{22}, \sigma_{12})^\top, (\sigma_{12}, \sigma_{22})^\top)$ where $\sigma_{22} = 1, 5$ and $10$ (left to right), as function of $\eta_1$ and $\sigma_{12}$. (b): Bivariate extremal coefficient of the truncated extremal-t process ($\theta_2^{\text{tET}}$, blue) and the extremal-t process ($\theta_2^{\text{ET}}$, red) with $\nu = 2, 5$ and $10$ (left to right), as a function of the correlation parameter $\rho$.

The bivariate extremal coefficient of the skewed Brown-Resnick model only partly depends on $\Sigma$ through $\gamma_{12}$, the semivariogram of the process $Y(s)$ defined in Theorem 2. Opting for a parametrisation based on semivariogram functions yields greater interpretability of the model parameters and allows for asymptotic independence in the long range when considering an unbounded variogram. Indeed, for some spatial lag $h$, taking the power-law semivariogram with range $\lambda > 0$ and smoothness $\vartheta \in (0, 2]$, i.e. $\gamma(h) = (h/\lambda)^\vartheta$, implies that, as $h \to \infty$, $\eta \equiv \eta(h) \to 0$ and therefore $\xi = \xi(h) \to 0$, yielding $\theta_2 \to 2$ by (18).

Estimating the $D$-dimensional parameter vector $\eta$ under the constraint $\eta^\top \mathbf{1} = 0$ is a task that we simplify by representing $\eta$ through spline functions. Let $\eta_i = \sum_{j=1}^{J} b_j K_j(s_i)$, $i = 1, \ldots, D$, where $K_j(s_i)$ is the $j$-th basis function evaluated at $s_i = (s_{i,1}, s_{i,2}) \in \mathcal{S}$, and $b_j$ is the coefficient for the $j$-th basis. Imposing $\sum_{i=1}^{D} K_j(s_i) = 0$ automatically fulfills the sum-to-zero constraint for $\eta$. In the remainder of this paper, $K_j(\cdot)$ denotes a Gaussian kernel centered at location $s_j^\star$, $j = 1, \ldots, J$, i.e.,

$$K_j(s_i) = \frac{K_j^\star(s_i)}{\sqrt{\sum_{k=1}^{D} K_j^\star(s_k)^2}}, \quad \text{where} \quad K_j^\star(s_i) = \phi\left(\frac{\|s_i - s_j^\star\|}{2\sigma_B}\right) - \frac{1}{D}\sum_{k=1}^{D} \phi\left(\frac{\|s_k - s_j^\star\|}{2\sigma_B}\right), \sigma_B > 0. \quad (20)$$

Figure 2 displays $\theta_2^{\text{sBR}}$ between locations over a $[0, 32] \times [0, 32]$ grid and a reference point shown as a red star, considering $\eta_i = \sum_{j=1}^{2} b_j K_j(s_i) + 0.1 \text{sgn}(s_{2,i} \geq 16)$ with $J = 2$ kernel centers (black dots in Figure 2), $s_1^\star = (8, 8)$ and $s_2^\star = (24, 24)$, and $\sigma_B = 64$. Note that $\text{sgn}(\cdot)$ denotes the sign function, which in the last term in $\eta_i$ serves as background noise to ensure numerical identifiability of $\xi = \frac{\Sigma \eta}{\sqrt{1 + \eta^\top \Sigma \eta}}$. This is because multiplying $\eta$ by a positive constant $c$, will not have a numerically significant change on $\xi$ as the $\eta^\top \Sigma \eta$ term will dominate the denominator. Focusing on a single reference point (star), i.e. on either the four left or right panels of Figure 2, we observe that varying the $\eta$ vector through $(b_1, b_2)$ produces a broad range of dependence structures. Similarly, for fixed parameter values $(b_1, b_2)$, different dependence behaviours are observed depending on the reference point, highlighting an important feature of the skewed Brown-Resnick process: the ability to model non-stationarity and local anisotropy. This important property, exhibited by numerous real-world environmental extreme phenomena, is due to the fact that (18) not only depends on the spatial lag between two locations, but also on the corresponding components of the $\boldsymbol{\xi}$ parameter vector. A similar observation was made by Beranger et al. (2017) about the extremal skew-$t$ model.

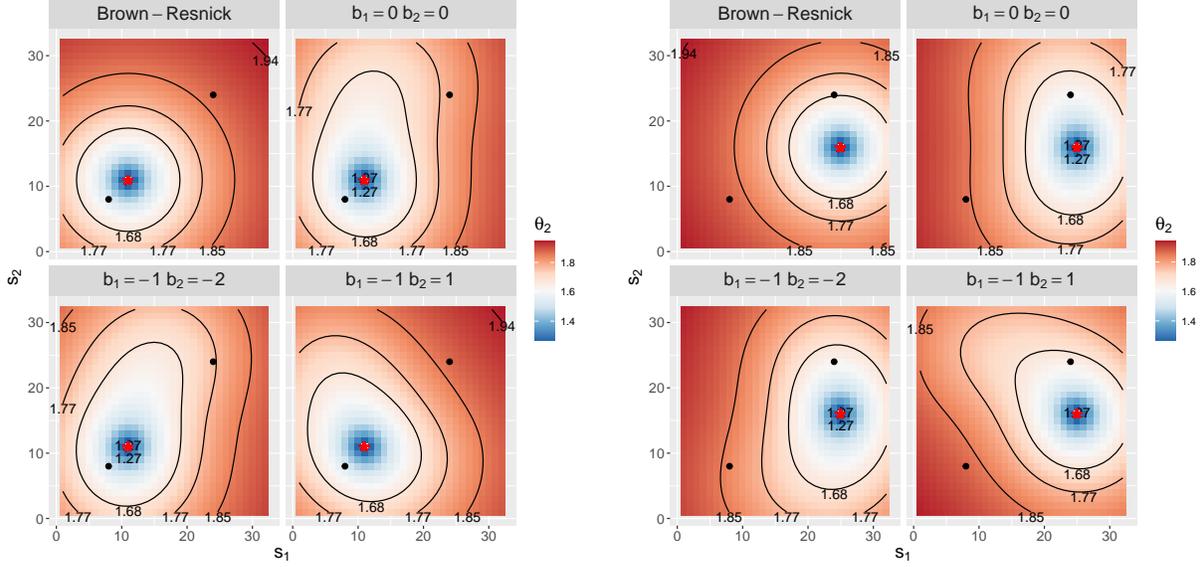

Figure 2: Heatmaps and contour levels (solid black lines) of the bivariate extremal coefficient for the Brown-Resnick model ($\eta_i = 0, i = 1, \ldots, D$) and skewed Brown-Resnick model where $\eta_i = \sum_{j=1}^{2} b_j K_j(s_i) + 0.1 \text{sgn}(s_{2,i} \geq 16), i = 1, \ldots, D$ using $(b_1, b_2) = (0,0), (-1,-2)$ and $(-1,-1)$. Black dots denote the kernel centres $s_1^\star, s_2^\star$. A red star indicates the reference point from which each bivariate extremal coefficient is taken with respect to.

# 5 Simulation experiments

## 5.1 Simulation algorithms

Following the approach of Dombry et al. (2016), Algorithms 1 & 2 (see Supplementary Material Sections C.4 and D.4) describe how to randomly generate data from the skewed Brown-Resnick and truncated extremal-$t$ models. Simulating from a $r$-Pareto process involves drawing from a random vector $\boldsymbol{Y}$ with distribution $P_{s_i}(\cdot) = \mathrm{E}_{\boldsymbol{W}}[\mathbb{1}_{\{\boldsymbol{W}/W(s_i) \in \cdot\}} W(s_i)]$ at a uniformly distributed random location $s_i$, where $\boldsymbol{W}$ is as in (1), implying that exact simulation algorithms for max-stable models can be leveraged (Dombry et al., 2016, 2024).

When the risk functional is the $L_1$ norm, i.e., $r_0(\cdot) = \|\cdot\|_1$, $r$-Pareto samples can be simply obtained as $\tilde{R} \boldsymbol{Y}/\|\boldsymbol{Y}\|_1$, where $\tilde{R}$ follows a unit Pareto distribution. For another risk functional $r_1(\cdot)$, let $M$ be such that $r_1(\cdot) \leq M r_0(\cdot)$. Samples for the $r$-Pareto process with risk functional $r_1(\cdot)$ can then be obtained by sampling $\tilde{Z}$ from the process with risk functional $r_0(\cdot)$, and accepting a sample as $\tilde{Z}/M$ when $r_1(\tilde{Z}) \geq M$ (Dombry et al., 2024, Proposition 2). The constant $M = 1$ when $r_1 = \|\cdot\|_\infty$ and $M = D^{p-2}$ when $r_1 = \|\cdot\|_p$, $p > 1$.

Larger $M$ implies lower acceptance probabilities. However, note that for a convex risk functional $r(\cdot)$, e.g., $L_p$ norm, $p \geq 1$, the support space $\mathcal{A}_r \in \mathbb{R}_+^D$ has complement $\mathcal{A}_r^c = \{x \in \mathbb{R}_+^D : r(x) \leq 1\}$, which is a convex set. Hence $\mathcal{A}_r \subset \mathcal{A}_{r_0}$. Following from this, an efficient simulation algorithm for a $r$-Pareto process with any convex risk functional $r(\cdot)$ is given below (see Supplemental Material E.1 for a proof).

**Theorem 4.** *Let $\tilde{Z}^{(r_0)}$ be a $r$-Pareto process at $D$ locations with risk functional $r_0(\cdot) = \|\cdot\|_1$. Let $r(\cdot)$ be any convex risk functional, homogeneous of order 1, such that $r(\mathbf{0}) = 0$ and $r(e_i) = 1/c_i > 0, i = 1, \ldots, D$, where $e_i$ are standard basis vectors of dimension $D$. The r-Pareto process with risk functional $r(\cdot)$ can be simulated as $\tilde{Z}^{(r)} \overset{d}{=} c_0 \tilde{Z}^{(r_0)} \mid r\left(\tilde{Z}^{(r_0)}\right) > 1/c_0$, where $c_0 = \min\{\min_{i=1}^D c_i, 1\}$.*

For example, considering $r(\cdot) = \|\cdot\|_p$, $p > 1$ gives $1/c_0 = 1$ instead of $M = D^{p-2}$ in Dombry et al. (2024), while $r(x) = \sum_{i=1}^D m_i x_i, m_i \in [0, \infty], i = 1\ldots, D$, a linear convex combination, produces $c_0 = \min\{\min_{i=1}^D 1/m_i, 1\}$. This algorithm is simpler than the one proposed by Dombry et al. (2024, Proposition 1) when $r(\cdot)$ is a linear convex combination. Table 1 shows the acceptance rate of both algorithms, when generating samples for a Brown-Resnick $r$-Pareto process with a $L_p$ risk functional, $p = 2, 3, 5, 10$, for varying dimension $D$. As $p$ increases, the acceptance rate decreases quickly to zero using method in Dombry et al. (2024), especially when $D$ is large. In contrast, our method achieves a comparable or higher acceptance rate in all cases, particularly for $p > 2$. However the approach of Dombry et al. (2024) is more general and can be used when $r(\cdot)$ is not convex, but homogeneous of order 1 (e.g., $r$ returns any order statistics excluding the maximum), once $M$ has been obtained. In the following simulations, Theorem 4 is used to randomly generate data, and is also leveraged in our proposed inference methodology (see Section 3).

Table 1: Percentage of samples from $10^5$ Brown-Resnick $r$-Pareto processes replicates with $L_1$ risk functional that fall into the acceptance region for $r(\cdot)$ being $L_p, p = 2, 3, 5, 10$, using the methods in Dombry et al. (2024) (left number in each cell) and Theorem 4 (right number). Dimension is $D = 4, 16, 64, 100$ and dependence structure is defined via a power-law semivariogram $\gamma(h) = h/\lambda$, $\lambda = 2$, on the grid $[1, \sqrt{D}]^2$.

| $L_p/D$ | $D = 4$ | $D = 16$ | $D = 64$ | $D = 100$ |
|---|---|---|---|---|
| $p = 2$ | 59/59 | 37/37 | 25/25 | 22/22 |
| $p = 3$ | 13/52 | 1.80/29 | 0.27/18 | 0.19/16 |
| $p = 5$ | 0.76/49 | >0.01/26 | 0.00/16 | 0.00/14 |
| $p = 10$ | >0.01/47 | 0.00/25 | 0.00/15 | 0.00/13 |

## 5.2 Parameter estimation

### 5.2.1 Spectral vs pairwise composite likelihoods

An initial experiment on the Brown-Resnick model (Supplementary Material F) indicates that, unlike the composite likelihood, the spectral likelihood produces biased estimates especially with weaker dependence, supporting the findings from Engelke et al. (2015); Huser et al. (2016). However, the pairwise likelihood is around 476 times slower to implement.

For our new models, composite likelihood is not directly applicable for truncated extremal-$t$ processes as the marginal distribution of a multivariate truncated normal is not truncated normal, and the actual density requires a costly numerical integration for each composite likelihood term (Horrace, 2005). We therefore focus on the skewed Brown-Resnick model defined on a $15 \times 15$ grid ($D = 225$) with parameters $\lambda = 3$, $\vartheta = 1$, and $(b_1, b_2) = (-1, -2)$, where the centres of the kernels are $s_1^\star = (4, 4)$ and $s_2^\star = (11, 11)$, and consider samples of size $n = 500$. The threshold for the spectral likelihood (7) is set at $u = 30 \times D$ and the composite likelihood weights in (6) are set to 0/1 to only include the 1,202 closest pairs, mitigating computational complexity. Violin plots of the estimates (for 300 replicate analyses) are displayed in Figure 3, demonstrating almost unbiased estimates for $b_1$ and $b_2$ and small relative bias (~10%) for $\lambda$ and $\vartheta$ when the spectral likelihood (red) is applied. In contrast, the pairwise likelihood estimates (blue) for $\lambda$ and $\nu$ are unbiased, but those for $b_1$ and $b_2$ are biased and almost unreliable, which may be attributed to the inability of this approach to capture the locally anisotropic dependence structure of the data. The spectral likelihood is about 68 times faster than the composite likelihood, taking 124 seconds on a 16-core machine (as for all computations in this manuscript) and considering multiple initial values for the skewness parameter to avoid local maxima.

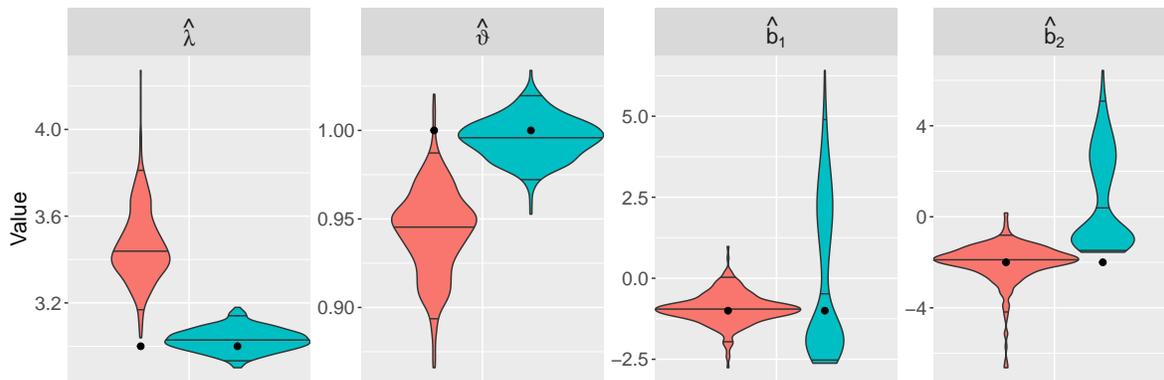

Figure 3: Violin plots for 300 estimates of $\boldsymbol{\theta} = (\lambda, \vartheta, b_1, b_2)$ for the skewed Brown-Resnick process with $\lambda = 3$, $\vartheta = 1$, and $(b_1, b_2) = (-1, -2)$ using the spectral likelihood (red; left) with threshold $u = 30 \times D$ and the pairwise likelihood (blue; right) with 1,202 pairs. Black dots indicate the true values.

### 5.2.2 Threshold selection for spectral likelihoods

First, we consider the skewed Brown-Resnick model with sample size $n = 2000$ and all 12 parameter combinations from $\lambda = 5, 10$, $\vartheta = 1, 1.5$, and $(b_1, b_2) = (0, 0), (-1, -2), (-1, 1)$. The grid size and kernel centres remain unchanged. The threshold in the spectral likelihood is set as $u = 100 \times D$, retaining the ~20 observations with largest radial component. Figure 4 displays violin plots for each parameter estimate and true parameter combination (indicated as cases 1 – 12), highlighting that all parameters were accurately identified, no matter the degree of dependence. This confirms that increasing $u$ reduces bias in the spectral likelihood since the Poisson approximation is better. Model fitting is computationally efficient, taking on average ~6 minutes per fit using 16 cores and setting multiple initial values for the skewness parameter.

We now generate $n = 2000$ samples from the truncated extremal-$t$ model defined on a 10×10 grid ($D = 100$) with 2 parameter combinations using $\lambda = 3, 5$, $\vartheta = 1$ and $\nu = 2$. Fixing $\nu = 2$, Figure 5 shows the results of spectral likelihood estimation where the threshold $u$ is set as the 98% (red), 95% (green), and 90% (blue) quantiles of the observed radial components. Retaining the 40 ($u = 98\%$) observations with largest radial components produces almost unbiased estimates. Lowering $u$ increases the bias but also increases computation time since more observations are included. As noted in Section 4.2, fitting the truncated extremal-$t$ model requires more computational effort than the skewed Brown-Resnick, taking on average ~15 minutes. Overall, a threshold should be as high as possible while retaining enough observations to control estimate variability.

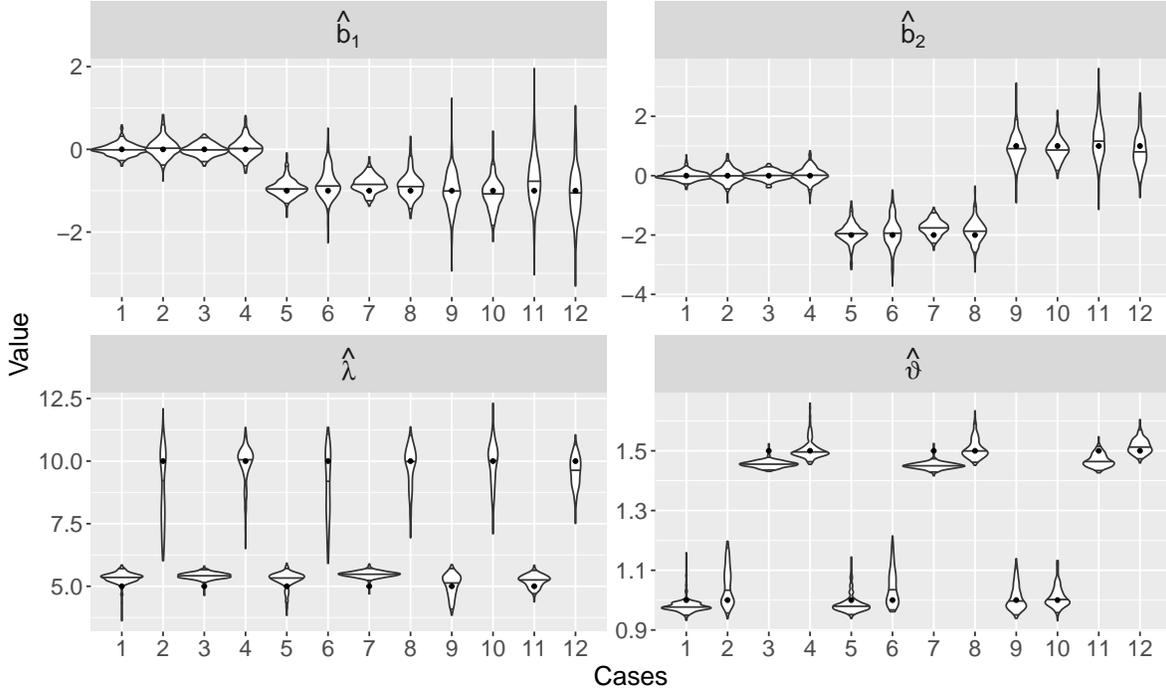

Figure 4: Violin plots for the 300 estimates of $\boldsymbol{\theta} = (\lambda, \vartheta, b_1, b_2)$ for the skewed Brown-Resnick process with all combinations of $\lambda = 5, 10$, $\vartheta = 1, 1.5$, and $(b_1, b_2) = (0, 0), (-1, -2), (-1, 1)$, using the spectral likelihood. Black dots indicate the true values.

### 5.2.3 Spectral likelihoods vs score matching

We now compare the performance of the spectral likelihood to the popular score matching approach (see de Fondeville and Davison, 2018) when fitting $r$-Pareto processes with different risk functionals. We sample $n = 2,000$ observations from the skewed Brown-Resnick $r$-Pareto process defined on a $15 \times 15$ grid with the same parameter combinations as in Figure 4 and taking the $L_1$ and $L_3$ norms as risk functionals. An observation is considered extreme when exceeding the 95% empirical quantile of $r(\boldsymbol{X}_1), \ldots, r(\boldsymbol{X}_n)$.

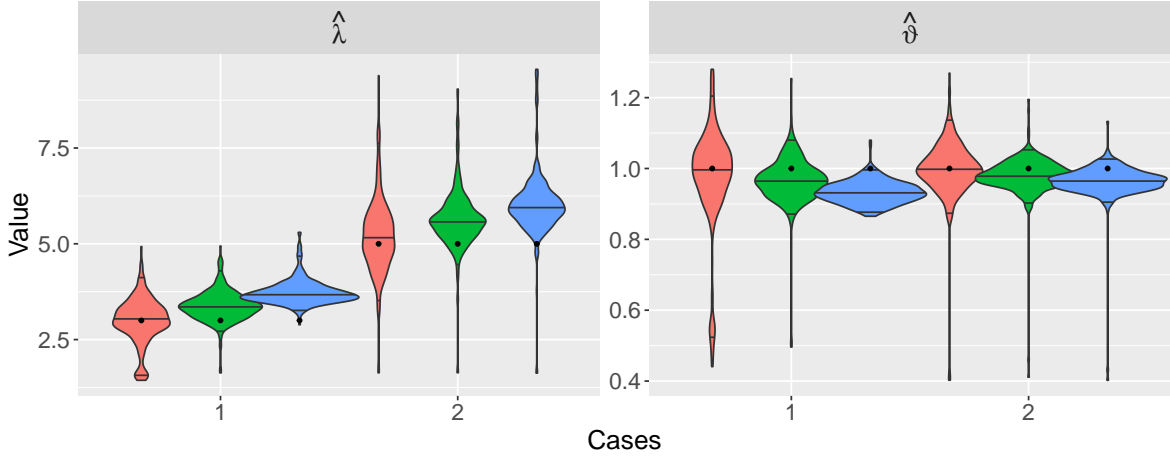

Figure 5: Violin plots for the spectral likelihood estimates of the truncated extremal-$t$ model with $\lambda = 3, 5$, $\vartheta = 1$, and $\nu = 2$ (fixed), and threshold $u$ set as the 98% (red), 95% (green) and 90% (blue) empirical quantiles of the radial component. Black dots indicate the true parameter values.

Estimates of $\vartheta$ and $b_1$ when the risk functional is the $L_3$ norm are shown in Figure 6. (Estimates for $\widehat{\lambda}$ and $\widehat{b}_2$ are reported in Figure 2 of Supplementary Material E.3.) The spectral likelihood consistently provides unbiased, low variability estimates (blue) for all parameters across all dependence structure regimes. In contrast, score matching produces unbiased but more variable estimates (red) and can become numerically unstable for the skewness parameters (e.g., cases 7–12). The spectral likelihood is also more computationally efficient, being ∼5 times faster than the score matching approach (141 versus 704 seconds on average using 3 CPU cores). Figures 3 and 4 of the Supplementary Material report similar results and conclusions when the risk functional is the $L_1$ norm and for the truncated extremal-$t$ $r$-Pareto process.

In conclusion, the spectral likelihood is a statistically and computationally efficient methodology that can infer the model parameters of max-stable and $r$-Pareto processes. For max-stable processes, a carefully selected threshold will control the bias in the estimates due to the Poisson approximation. There is no bias (or approximation) for $r$-Pareto processes. The computational gains are significant, which enables the tackling of higher dimensional problems.

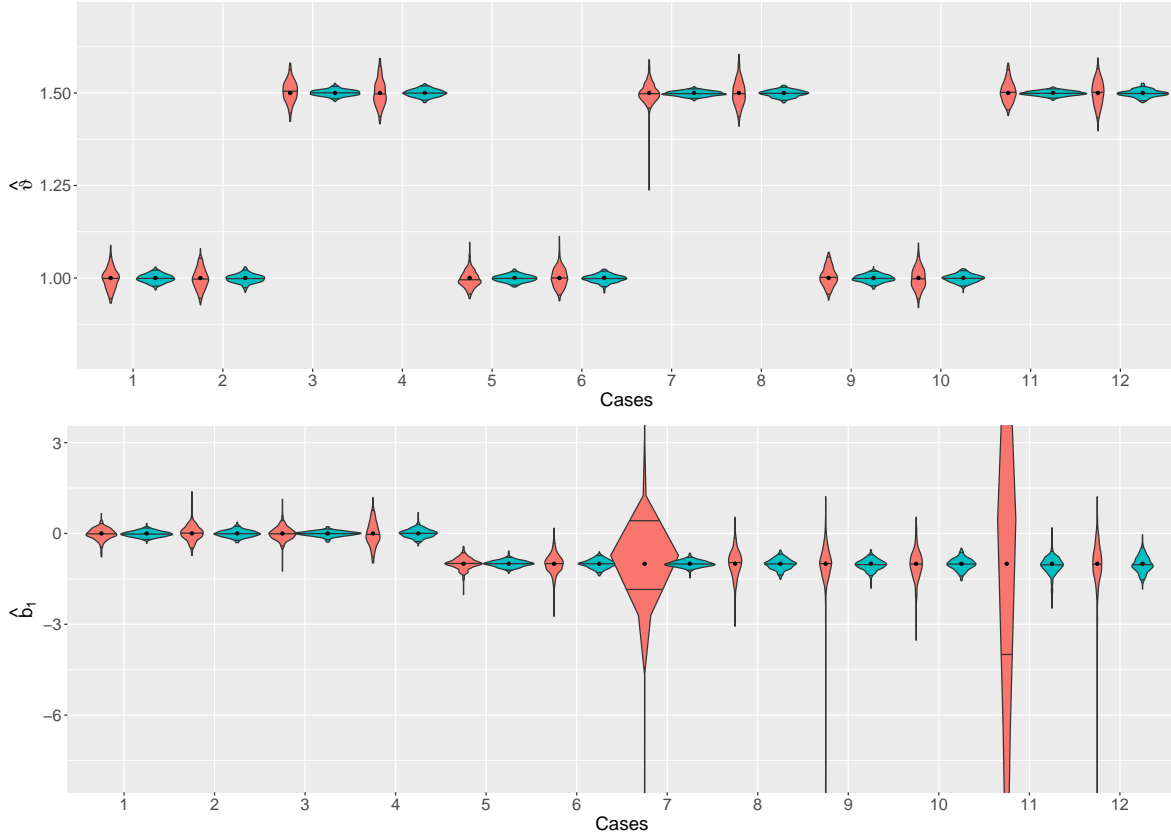

Figure 6: Violin plots for score matching (red) and spectral likelihood (blue) estimates of $\vartheta$ (top) and $b_1$ (bottom) for the skewed Brown-Resnick $r$-Pareto process with $L_3$ norm risk functional and the same parameter combinations as in Figure 4. Black dots indicate the parameter true values.

# 6 Practical applications

## 6.1 Analysis of extreme rainfall over Florida

We analyse the spatial dependence structure rainfall across the Tampa Bay area, Florida, during the wet season (June–September). The data comprise radar measurements recorded at 15 minute intervals between 1995–2019, over a regular 2km grid. The dataset is made of $n = 139,881$ radar images with $4,449$ spatial observations in each image. de Fondeville and Davison (2018) fitted a $r$-Pareto process based on Brown-Resnick model to these data restricted to the 1999–2004 period and $3,600$ locations. To motivate the need for a broader class of models, we fit the skewed Brown-Resnick model with spline representation of $\eta$ as in Section 4.3 with four kernels (see black dots in Figure 7) to the full dataset. As a pre-processing step, the data are marginally transformed to unit Pareto scale using the empirical distribution function, while zero values are ignored and treated as missing. We adopt an anisotropic semivariogram function

$$\gamma(\boldsymbol{h}) = \left(\frac{\|V\boldsymbol{h}\|_2}{\lambda}\right)^{\vartheta}, \; \vartheta \in (0, 2], \; \lambda > 0, \tag{21}$$

where $\boldsymbol{h}$ is the difference vector between coordinates of two locations and

$$V = \begin{pmatrix} \cos \zeta & -\sin \zeta \\ m \sin \zeta & m \cos \zeta \end{pmatrix}, \; m > 0, \; \zeta \in (-\tfrac{\pi}{4}, \tfrac{\pi}{4})$$

is a matrix used to capture the global anisotropy in the dependence structure. The $r$-Pareto processes are defined using the $L_1$ and $L_\infty$ norms as the risk functionals, motivated by the fact that the $L_\infty$ norm defines extremes events as locally intense rainfall events at any location within the region, whereas the $L_1$ norm selects events with high cumulative rainfall over the whole region. Although these types of events are of different nature, both can lead to flooding and both have recently occurred in Florida. For example, a plume of moisture caused heavy rainfall in southern Florida on June 11, 2024 generating high risks of localised flash flooding (Doermann, 2024), whereas hurricanes Irma and Milton (September 2017 and October 2024) lasted several days and touched large parts of the state (Cangialosi et al., 2021; Beven et al., 2025). To choose the thresholds, a univariate generalized Pareto distribution is fitted to the values of the corresponding risk functional evaluated at each observations and the level for which the estimated shape parameter becomes stable is selected. This produces a threshold set at the 99.95% empirical quantile of $\{r(\boldsymbol{X}_i)\}$ for both risk functionals, leaving 70 observations for model fitting.

Table 2 reports parameter and standard error estimates for both skew (sBR) and standard (BR) Brown-Resnick models and risk functionals ($L_1$ and $L_\infty$). The range parameter ($\lambda$) estimates are given in km/unit. For the Brown-Resnick models, spectral likelihood (BR) and score-matching (BR$^{sm}$) produce broadly consistent parameter estimates for both risk functions, although score matching produces notably smaller estimates of the range parameter $\lambda$. Figure 5 of the Supplementary Material shows the extremal coefficients of the Brown-Resnick model for both inference approaches, which strongly suggests the range parameter estimates using score matching are un-derestimated. Similar to Section 5.2, using the spectral likelihood with $L_1$ and $L_\infty$ risk functions is is respectively 80% and 18% faster than score matching. Inference for the sBR requires greater computation, reflecting its 4 additional parameters.

| Model | $r(\cdot)$ | $\widehat{\lambda}$ | $\widehat{\vartheta}$ | $\widehat{\zeta}$ | $\widehat{m}$ | $\widehat{b}_1$ | $\widehat{b}_2$ | $\widehat{b}_3$ | $\widehat{b}_4$ | Time (hrs) | AIC |
|---|---|---|---|---|---|---|---|---|---|---|---|
| BR$^{sm}$ | $L_1$ | 2.80 | 1.40 | -0.12 | 1.06 | - | - | - | - | 2.05 | - |
|  |  | (0.25) | (0.09) | (1.25) | (0.07) | - | - | - | - |  |  |
| BR | $L_1$ | 5.91 | 1.16 | -0.73 | 0.96 | - | - | - | - | 1.14 | 2485.88 |
|  |  | (1.71) | (0.11) | (1.21) | (0.07) | - | - | - | - |  |  |
| BR$^{sm}$ | $L_\infty$ | 3.49 | 1.71 | -0.04 | 1.02 | - | - | - | - | 1.90 | - |
|  |  | (0.22) | (0.04) | (1.83) | (0.05) | - | - | - | - |  |  |
| BR | $L_\infty$ | 6.30 | 1.37 | 0.40 | 0.95 | - | - | - | - | 1.70 | 2526.61 |
|  |  | (0.39) | (0.03) | (0.36) | (0.04) | - | - | - | - |  |  |
| sBR | $L_1$ | 5.81 | 1.19 | 0.73 | 1.04 | 11.39 | 8.58 | 23.09 | 19.89 | 4.20 | 2482.39 |
|  |  | (1.74) | (0.11) | (2.83) | (0.15) | (27.15) | (21.37) | (56.21) | (47.90) |  |  |
| sBR | $L_\infty$ | 5.96 | 1.41 | 0.46 | 0.96 | 0.16 | 0.15 | 0.06 | 0.07 | 4.94 | 2510.44 |
|  |  | (0.35) | (0.03) | (0.28) | (0.04) | (1.22) | (1.22) | (0.52) | (0.48) |  |  |

Table 2: Spectral likelihood parameter estimates (with jackknife estimates of standard errors) for the Brown-Resnick (BR) and skewed Brown-Resnick (sBR) $r$-Pareto models with $L_1$ and $L_\infty$ norm risk functionals. The superscript *sm* indicates inference performed using score matching.

Focusing on the spectral likelihood results, there is no evidence against $m = 1$, implying that there is no global anisotropy in the dependence structure, a finding consistent with de Fondeville and Davison (2018). The standard errors of $b_1, \ldots, b_4$ are relatively large, and although confidence intervals include zero, this does not imply there is no evidence these parameters are non-zero. Indeed, unlike in Section 5.2, no background noise has been added since only $\boldsymbol{\xi}$ needs to be estimated (instead of $\boldsymbol{\eta}$) and the identifiability of $b_1, \ldots, b_4$ is not a concern but reflected in the jackknife estimates of the standard errors. The greater flexibility offered by the skewed Brown-Resnick is demonstrated by lower AIC values for both risk functions. Preference for the sBR model indicates that a local anisotropic dependence structure is detected.

Figure 7 provides maps of empirical bivariate extremal coefficients (shading) with respect to two different reference points (south-west, north-west) and the two risk functionals ($L_1$ top panels; $L_\infty$ bottom panels). We note that the empirical extremal coefficient patterns appear to vary with the reference location, indicating a non-stationary dependence structure. The ability of the sBR model (and the inability of the BR model) to capture these local dependence features can be seen in the dependence contours of the BR (solid lines) and sBR (dashed lines) models. Comparing insights drawn between risk functionals, with a smaller range of dependence the $L_\infty$ norm (bottom plots) exhibits a more localised extremal dependence structure, whereas the $L_1$ norm (top plots) exhibits dependence that is more persistent with distance. This observation is consistent with the interpretation of the extreme events selected by these risk functionals: the $L_1$ norm considers cumulative rainfall over the whole region, and the $L_\infty$ norm identifies events more consistent with local area rainfall.

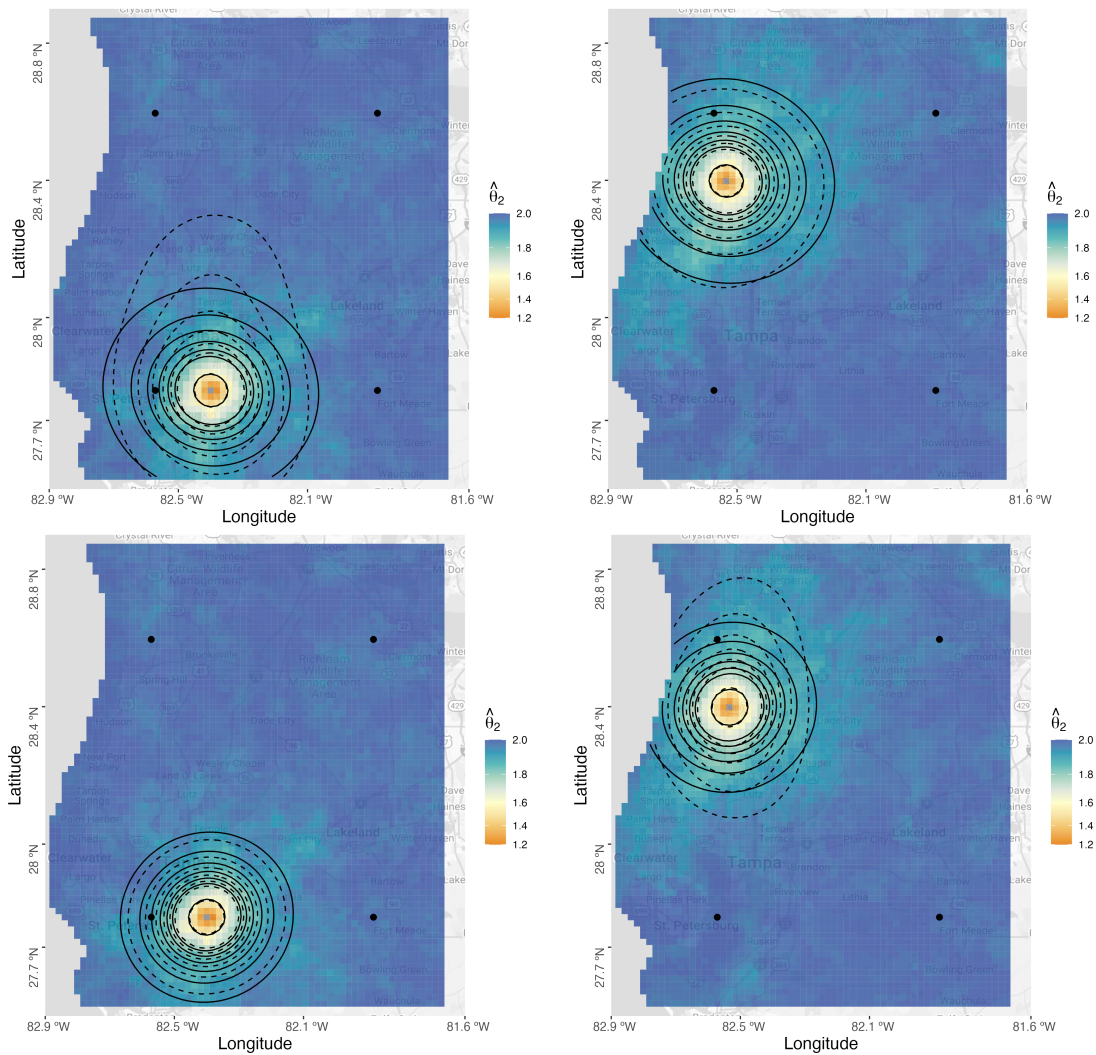

Figure 7: Maps of bivariate empirical extremal coefficients (shading) with respect to two different reference points (columns), and contours of the extremal coefficient of the fitted sBR (dashed line) and BR (solid line) $r$-Pareto models on the Florida rainfall data. Top and bottom rows use the $L_1$ and $L_\infty$ norm risk functional, respectively. Black dots denote the kernel centres used in the sBR model.

## 6.2 Analysis of sea surface temperature over the Red Sea

We analyse the spatial dependence structure of the sea surface temperature (in °C) over the Red Sea. An exploratory analysis by Hazra and Huser (2021) demonstrated the necessity for a model that captures asymptotic dependence as well non-stationarity. The data comprise 31 annual maxima recorded between 1985 and 2015 at $D = 1,043$ gridded locations. Following Huser et al. (2023) the annual maxima are marginally standardized to unit Fréchet scale. The BR and sBR max-stable process models are fitted using the spectral likelihood with threshold $u = 10 \times D$, where 10 is approximately the 90% quantile of the unit Fréchet distribution, leaving 9 observations for model fitting. The spline representation of $\eta$ uses 5 kernel centres along the Red Sea's North-South axis (see black dots in Figure 8). and as with the previous analysis, no background noise is added.

We contrast model fitting with the spectral likelihood, composite likelihood and Vecchia approximations. The latter was proposed by Huser et al. (2023) for fitting max-stable process models, who demonstrated that it could outperform composite likelihoods when using the Brown-Resnick model with fixed smoothness parameter $\vartheta$. We implement this approach with the anisotropic semi-variogram (21) and directly estimating $\vartheta$. The composite likelihood uses all pairs of first-order (immediate N, S, E, W) neighbours on the grid, yielding 2,046 pairs. The parameter estimation results are shown in Table 3, with only spectral likelihoods being used for the sBR model.

| Model | $\widehat{\lambda}$ | $\widehat{\vartheta}$ | $\widehat{\zeta}$ | $\widehat{m}$ | $\widehat{b}_1$ | $\widehat{b}_2$ | $\widehat{b}_3$ | $\widehat{b}_4$ | $\widehat{b}_5$ | Time (hrs) | AIC |
|---|---|---|---|---|---|---|---|---|---|---|---|
| BR$^c$ | 58.61 | 1.35 | -0.54 | 0.76 | - | - | - | - | - | 0.93 | - |
|  | (14.79) | (0.18) | (0.05) | (0.04) | - | - | - | - | - |  |  |
| BR$^v$ | 108.91 | 1.16 | 0.79 | 1.21 | - | - | - | - | - | 1.41 | - |
|  | (12.20) | (0.09) | (0.10) | (0.06) | - | - | - | - | - |  |  |
| BR | 64.78 | 1.16 | 0.79 | 1.14 | - | - | - | - | - | 0.17 | 1871.82 |
|  | (7.06) | (0.10) | (0.16) | (0.05) | - | - | - | - | - |  |  |
| sBR | 51.36 | 1.33 | 0.79 | 1.12 | -13.49 | -8.31 | 2.05 | 12.85 | 15.07 | 0.65 | 1843.46 |
|  | (4.75) | (0.09) | (0.28) | (0.07) | (15.48) | (6.76) | (9.32) | (7.58) | (11.47) |  |  |

Table 3: Parameter estimates (with jackknife estimates of the standard errors) for the Brown-Resnick (BR) and skewed Brown-Resnick (sBR) max-stable process models on the Red Sea data. Model name with superscript $v$ or $c$ denotes inference using Vecchia or composite likelihood approximations, respectively. No superscript indicates use of the spectral likelihood.

Focusing on the Brown-Resnick parameter estimates, Table 3 indicates broadly comparable statistical efficiency across all methods and Vecchia estimates in agreement with those observed by Huser et al. (2023). By displaying the fitted extremal coefficients, Figure 6 of the Supplementary Material illustrates that even though the pairwise likelihood estimates (BR$^c$) reported in Table 3 seem slightly different to those from the other methods, they lead to a similar dependence structure. In terms of computational efficiency the spectral likelihood is respectively ∼ 4.5 and ∼ 7.3 times faster than the composite and Vecchia approaches. Fitting the skewed Brown-Resnick model with the spectral likelihood yields a smaller estimate of $\lambda$ and a larger estimate of $\vartheta$ than for the Brown-Resnick. This is somewhat expected as the fitted skewed Brown-Resnick model better accounts for the dependence strength and structure than the Brown-Resnick model. Even with the additional parameters, the AIC favours the skewed Brown-Resnick model over the regular Brown-Resnick model, indicating improved modelling of spatial dependence. We also note that fitting the sBR model using the spectral likelihood is faster than fitting the BR using the Vecchia or composite likelihood approaches.

To assess the goodness of fit and visualise the differences in dependence structures from the fitted models, Figure 8 shows maps of the bivariate extremal coefficient for two reference locations. Both empirical maps (shading) display some degree of non-stationarity in the dependence structure, in agreement with Hazra and Huser (2021), with the contours of the extremal coefficient obtained from the skewed Brown-Resnick model (dashed lines) being able to better capture this phenomena. In particular, the skewed Brown-Resnick model is able to simultaneously characterise the differing behaviour observed in the north and south of the Red Sea.

# 7 Discussion

In this paper, we have established the condition ensuring the intensity function of a max-stable process only places mass on $\Omega^\circ$, the interior of its domain. This condition directly implies that there are no discontinuities in the associated exponent measure, greatly simplifying the evaluation of the finite dimensional density of the corresponding $r$-Pareto process. As a result, we were able to demonstrate that for models possessing the above property, likelihood-based inference can be successfully implemented via the spectral likelihood (7). This both opens up viable solutions to the likelihood intractability of max-stable models, and also creates opportunities for the development of new classes of $r$-Pareto models that can better describe real-world processes. Indeed, out of the two primary max-stable models in current use, namely the Brown-Resnick and extremal-$t$, only the former fulfills the continuity condition and is therefore, almost exclusively, the only $r$-Pareto model used in the literature.

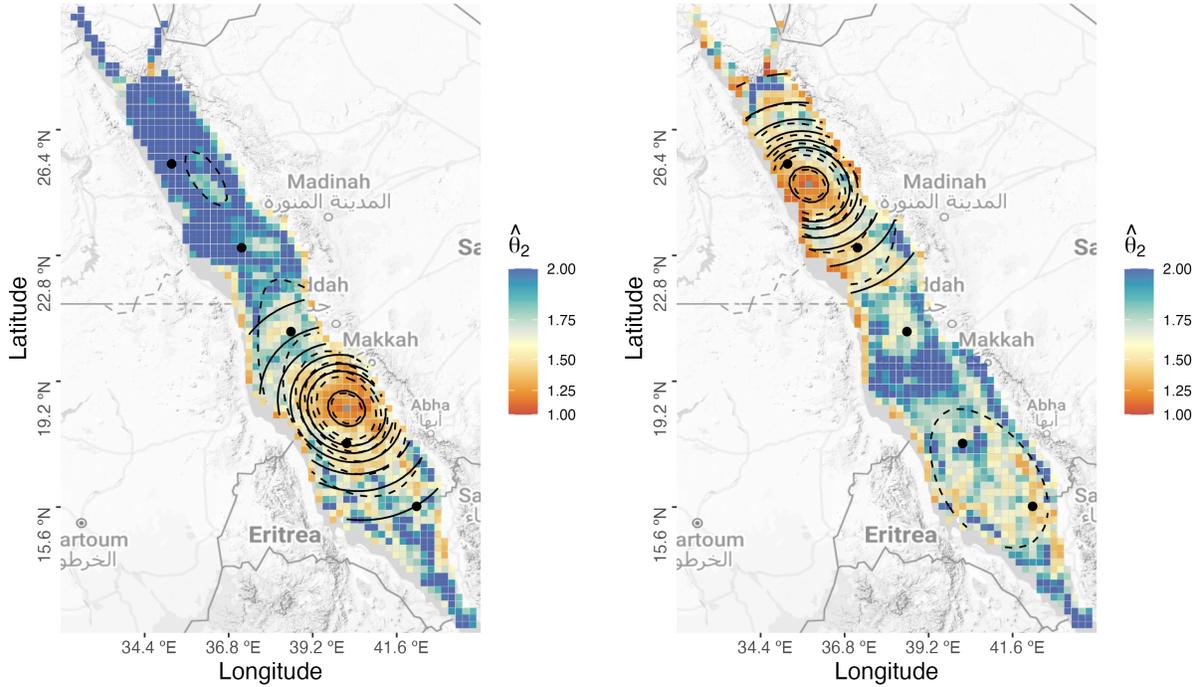

Figure 8: Maps of the bivariate empirical extremal coefficient (shading) with respect to two reference locations (left and right panels) with contours curves of the estimated extremal coefficients from the Brown-Resnick (solid) and skewed Brown-Resnick (dashed) models. Black dots indicate the kernel centres used in the skewed Brown-Resnick model.

As a first step to address this, we derived two new models for spatial extremes: the skewed Brown-Resnick model, which offers greater flexibility than the Brown-Resnick model by allowing for non-stationarity and local anisotropy; and the truncated extremal-$t$ model, which mimics the extremal-$t$ while ensuring the continuity condition holds. Exact simulation algorithms for these new max-stable models can leverage existing simulation techniques, whereas for the associated $r$-Pareto processes we have introduced a more efficient simulation algorithm than the current state-of-the-art. Following the reasoning of the latter, we have demonstrated that the inferential intractability issues for $r$-Pareto processes with any risk functional can be bypassed by simply considering the likelihood of a $r$-Pareto process with $L_1$ risk functional.

Using these new models, we have demonstrated the need for better, more flexible models for modelling both threshold exceedances and annual maxima data, by providing new insights about the non-stationary dependence structure of extreme precipitation and sea surface temperature levels over previous analyses (de Fondeville and Davison, 2018; Huser et al., 2023). The proposed methodology is computationally more efficient than state-of-the-art techniques, such as score matching (de Fondeville and Davison, 2018) and Vecchia approximation (Huser et al., 2023), thereby enabling the analysis of much higher dimensional data.

The new models introduced here are one step towards building spatial extreme models that are well-suited for real-world problems. They also open new opportunities in related areas, such as expanding the class of extremal graphical models with flexible dependence structures, as currently the only practical models for graphical extremes are constructed from the restrictive Brown-Resnick or Hüsler-Reiss models (Engelke and Hitz, 2020; Hentschel et al., 2024; Wan and Zhou, 2023).

# Supplemental Material

This contains proofs of all theorems and propositions, and exact simulation algorithms for the skewed Brown-Resnick and truncated extremal-t models. Code for the simulations and applications is available at https://github.com/PangChung/InteriorExtremes.

# Acknowledgements

This work was supported by the Australian Research Council (ARC) through the Discovery Project Scheme (DP220103269) and the Centre of Excellence for Climate Extremes (CE170100023).

# Supplemental Material for "Fast and flexible inference for spatial extremes"

Peng Zhong [*], Scott A. Sisson and Boris Béranger

May 30, 2025

In the following, equation numbers with prefix, $S$, refer to equations in the supplemental material, otherwise refer to equations in the main manuscript.

## A  Proof of Theorem 1

For any $k \in \{1, \ldots, D-1\}$, we have

$$\lim_{\boldsymbol{x}_{\overline{B}_k} \downarrow \boldsymbol{0}_{D-k}} -V_{B_k}(\boldsymbol{x}) = \int_0^\infty r^k f_{\boldsymbol{W}_{B_k}}(r\boldsymbol{x}_{B_k}) \lim_{\boldsymbol{x}_{\overline{B}_k} \downarrow \boldsymbol{0}_{D-k}} \Pr(\boldsymbol{W}_{\overline{B}_k} \in [\boldsymbol{0}, r\boldsymbol{x}_{\overline{B}_k}] \mid \boldsymbol{W}_{B_k} = r\boldsymbol{x}_{B_k}) \mathrm{d}r$$

$$= \int_0^\infty r^k f_{\boldsymbol{W}_{B_k}}(r\boldsymbol{x}_{B_k}) \Pr(\boldsymbol{W}_{\overline{B}_k} = \boldsymbol{0}_{D-k} \mid \boldsymbol{W}_{B_k} = r\boldsymbol{x}_{B_k}) \mathrm{d}r,$$

which implies that if $\Pr(\boldsymbol{W}_{\overline{B}_k} = \boldsymbol{0}_{D-k} \mid \boldsymbol{W}_{B_k} = r\boldsymbol{x}_{B_k}) = 0$ then $\lim_{\boldsymbol{x}_{\overline{B}_k} \downarrow \boldsymbol{0}_{D-k}} -V_{1:k}(\boldsymbol{x}) = 0$. Moreover, if the intensity function on $\partial \Omega$ is zero everywhere, then we must have

$$\lim_{\boldsymbol{x}_{\overline{B}_k} \downarrow \boldsymbol{0}_{D-k}} -V_{B_k}(\boldsymbol{x}) = \int_0^\infty r^k f_{\boldsymbol{W}_{B_k}}(r\boldsymbol{x}_{B_k}) \Pr(\boldsymbol{W}_{\overline{B}_k} = \boldsymbol{0}_{D-k} \mid \boldsymbol{W}_{Bk} = r\boldsymbol{x}_{B_k}) \mathrm{d}r = 0,$$

and since the integrand is non-negative, we conclude that the integrand must be zero almost everywhere. Therefore, we have $\Pr(\boldsymbol{W}_{\overline{B}_k} = \boldsymbol{0}_{D-k} \mid \boldsymbol{W}_{B_k} = r\boldsymbol{x}_{B_k}) = 0$ for any $r > 0$ and $\boldsymbol{x}_{B_k} > \boldsymbol{0}_k$ almost everywhere because $r^k$ is positive and $W$ is a non-negative continuous random process, which implies the condition (13) must hold.

---

[*]School of Mathematics and Statistics, UNSW Sydney, Australia.
{Peng.Zhong, Scott.Sisson, B.Beranger}@unsw.edu.au

# B  Useful distributions

**Definition 1** (Extended skew-normal distribution Azzalini (2013, Chapter 5)). *A random vector $Y = (Y_1, \ldots, Y_D) \in \mathbb{R}^D$ is said to follow the extended skew-normal distribution with location vector $\boldsymbol{\mu} \in \mathbb{R}^D$, $D \times D$ scale matrix $\Sigma$, slant parameter $\boldsymbol{\alpha} \in \mathbb{R}^D$ and extension parameter $\tau \in \mathbb{R}$ if its probability density function (pdf) can be written as*

$$\psi(x; \boldsymbol{\mu}, \Sigma, \boldsymbol{\alpha}, \tau) = \phi(x; \boldsymbol{\mu}, \Sigma) \frac{\Phi(\alpha_0 + \boldsymbol{\alpha}^\top \boldsymbol{\omega}^{-1}(x - \boldsymbol{\mu}); 0, 1)}{\Phi(\tau; 0, 1)},$$

*where $\alpha_0 = \tau \left(1 + \boldsymbol{\alpha}^\top \overline{\Sigma} \boldsymbol{\alpha}\right)^{1/2}$, $\overline{\Sigma} = \boldsymbol{\omega}^{-1} \Sigma \boldsymbol{\omega}^{-1}$, $\boldsymbol{\omega} = \sqrt{\text{diag}(\Sigma)}$. The corresponding cumulative distribution function (cdf) is given by*

$$\Psi(x; \mu, \Sigma, \alpha, \alpha_0) = \frac{\Phi\left(\begin{bmatrix} \boldsymbol{\omega}^{-1}(x - \boldsymbol{\mu}) \\ \tau \end{bmatrix}; \begin{bmatrix} 0 \\ 0 \end{bmatrix}, \begin{bmatrix} \overline{\Sigma} & -\frac{\overline{\Sigma}\boldsymbol{\alpha}}{(1+\boldsymbol{\alpha}^\top \overline{\Sigma} \boldsymbol{\alpha})^{1/2}} \\ -\frac{\boldsymbol{\alpha}^\top \overline{\Sigma}}{\left(1+\boldsymbol{\alpha}^\top \overline{\Sigma} \boldsymbol{\alpha}\right)^{1/2}} & 1 \end{bmatrix}\right)}{\Phi(\tau; 0, 1)}.$$

*In short, we can write $Y \sim ESN(\boldsymbol{\mu}, \Sigma, \boldsymbol{\alpha}, \alpha_0)$.*

**Definition 2** (Multivariate truncated normal distribution). *Let $Y$ be a D-dimensional Gaussian random vector with mean $\mu \in \mathbb{R}^D$ and covariance matrix $\Sigma$, with pdf and cdf represented by $\phi(\cdot; \mu, \Sigma)$ and $\Phi(\cdot; \mu, \Sigma)$. The random vector $\widehat{Y} = Y | Y \geq 0$ defines a truncated Gaussian random vector with pdf*

$$f_{\widehat{Y}}(y) = c^{-1} \phi(y; \mu, \Sigma), \quad y \in \mathbb{R}_+^D,$$

*where $c = \int_0^\infty \phi(y; \mu, \Sigma) dy$. In short, we can write $\widehat{Y} \sim \mathcal{N}_{[0,\infty]}(\mu, \Sigma)$.*

Note that while the conditional distributions of $\widehat{Y} \sim \mathcal{N}_{[0,\infty]}(\mu, \Sigma)$ remain within the truncated Gaussian family, the marginal distributions do not observe such a property (Horrace, 2005). Denote the set $A = [-\infty, y] = [-\infty, y_1] \times \cdots \times [-\infty, y_D]$ and $A_{B_k} = \cup_{b \in B_k} A_b$, where $A_b = [-\infty, y_1] \times \cdots \times [-\infty, y_{b-1}] \times [-\infty, 0] \times [-\infty, y_{b+1}] \times \cdots \times [-\infty, y_D]$. This implies that $[0, y] = A \cap \overline{A_{B_D}} = $

$A \cap_{b \in B_D} \overline{A}_b$, and by the inclusion-exclusion principle we have

$$\int_0^y \phi(x;\mu,\Sigma)\mathrm{d}x = \int_A \phi(x;\mu,\Sigma)\mathrm{d}x + \sum_{k=1}^D \sum_{B_k} (-1)^k \int_{A_{B_k}} \phi(x;\mu,\Sigma)\mathrm{d}x$$

$$= \Phi(y;\mu,\Sigma) + \sum_{k=1}^D \sum_{B_k} (-1)^k \Phi(y^{B_k};\mu,\Sigma) := \Phi^\circ(y;\mu,\Sigma), \qquad (S.1)$$

where $y^{B_k}$ corresponds to the vector $y$ where the $b$-th component, $b \in B_k$, is replaced by zero.

**Definition 3** (Cdf of the multivariate truncated normal distribution). *Let $\widehat{Y} \sim \mathcal{N}_{[0,\infty]}(\mu, \Sigma)$. The cdf of the truncated Gaussian is given by*

$$F_{\widehat{Y}}(y) = \tfrac{\Phi^\circ(y;\mu,\Sigma)}{\Phi^\circ(\infty;\mu,\Sigma)}, \quad y \in \mathbb{R}_+^D,$$

*where the denominator is given by* (S.1).

**Definition 4.** *Let $t_\nu(y;\mu,\Sigma)$ and $T_\nu(y;\mu,\Sigma)$ be the pdf and cdf of the multivariate t distribution with mean $\mu \in \mathbb{R}^D$, positive scale matrix $\Sigma$ and degree of freedom $\nu > 0$. Following a similar construction as in Definition 2, a random vector $Y$ is said to follow the multivariate truncated student-t distribution on $\mathbb{R}_+^D$ if its pdf can be written as*

$$\tfrac{t_\nu(y;\mu,\Sigma)}{T_\nu^\circ(\infty;\mu,\Sigma)}, \quad y \in \mathbb{R}_+^D.$$

*Additionally, its cdf is given by $T_\nu^\circ(y;\mu,\Sigma)/T_\nu^\circ(\infty;\mu,\Sigma)$. In short, we can write $Y \sim t_{[0,\infty]}(\mu, \Sigma, \nu)$.*

## C  Skew Brown-Resnick processes

### C.1  Proof of Theorem 2

Let $W = \exp\{Y - a\}$, where $Y$ is the random vector associated with the process $Y(s)$, i.e., $Y \sim \mathrm{ESN}(0, \Sigma, \alpha, 0)$ with scale matrix $\Sigma$ and slant vector $\alpha$, and $a = (a_1, \ldots, a_D)$, $a_k = \log \mathbb{E}\left[\exp\{Y_k\}\right]$, $k = 1, \ldots, D$.

By applying the first half of (2), we have

$$V(\boldsymbol{x}) = \sum_{k=1}^{D} \frac{1}{x_k} \int_{-\infty}^{\infty} \int_{-\infty}^{y_k + \boldsymbol{a}_{-k} - a_k \mathbf{1}_{D-1} + \log(\boldsymbol{x}_{-k}/x_k)} e^{y_k - a_k} \psi(\boldsymbol{y}; \mathbf{0}, \Sigma, \boldsymbol{\alpha}, 0) \mathrm{d}\boldsymbol{y}_{-k} \mathrm{d}y_k$$

$$= \sum_{k=1}^{D} \frac{\Phi(\tau_k; 0, 1)}{2} \frac{\exp\{\tilde{\omega}_k^2/2 - a_k\}}{x_k} \int_{-\infty}^{\infty} \int_{-\infty}^{y_k + \boldsymbol{a}_{-k} - a_k \mathbf{1}_{D-1} + \log(\boldsymbol{x}_{-k}/x_k)} \psi(\boldsymbol{y}; \Sigma_{:;k}, \Sigma, \boldsymbol{\alpha}, \tau_k) \mathrm{d}\boldsymbol{y}_{-k} \mathrm{d}y_k$$

(S.2)

where $\tau_k = \boldsymbol{\alpha}^\top \omega^{-1} \Sigma_{:;k} \left(1 + \boldsymbol{\alpha}^\top \overline{\Sigma} \boldsymbol{\alpha}\right)^{-1/2}$, $\overline{\Sigma} = \omega^{-1} \Sigma \omega^{-1}$, $\omega = \sqrt{\mathrm{diag}(\Sigma)}$ and $\tilde{\boldsymbol{\omega}} = (\tilde{\omega}_1, \ldots, \tilde{\omega}_D) = \omega \mathbf{1}$. The above integrals represent $\Pr[\boldsymbol{X}_{-k} - X_k \leq \boldsymbol{a}_{-k} - a_k \mathbf{1}_{D-1} + \log(\boldsymbol{x}_{-k}/x_k)]$, with $\boldsymbol{X} \sim \mathrm{ESN}(\Sigma_{:;k}, \Sigma, \boldsymbol{\alpha}, \tau_k)$. The random vector $\boldsymbol{X}_{-k} - X_k$ can be derived from $\boldsymbol{X}$ by applying the affine transformation $A_k \boldsymbol{X}$ where $A_k = (\boldsymbol{e}_1, \cdots, \boldsymbol{e}_{k-1}, -\mathbf{1}_{D-1}, \boldsymbol{e}_k, \ldots, \boldsymbol{e}_{D-1})$ and $\boldsymbol{e}_k, k = 1, \ldots, D-1$ are standard basis vectors of dimension $(D-1)$. From Azzalini (2013, Chapter 5) we have

$$\boldsymbol{X}_{-k} - X_k \sim \mathrm{ESN}\left(\boldsymbol{\mu}_k, \Sigma_k, \widehat{\boldsymbol{\alpha}}_k, \tau_k\right)$$

where $\boldsymbol{\mu}_k = A_k \Sigma_{:;k}, \Sigma_k = A_k \Sigma A_k^\top, \widehat{\boldsymbol{\alpha}}_k = (1 - \boldsymbol{\delta}^\top \omega A_k^\top \Sigma_k^{-1} A_k \omega \boldsymbol{\delta})^{-1/2} \widehat{\omega}_k \Sigma_k^{-1} A_k \omega \boldsymbol{\delta}, \boldsymbol{\delta} = \left(1 + \boldsymbol{\alpha}^\top \overline{\Sigma} \boldsymbol{\alpha}\right)^{-1/2} \overline{\Sigma} \boldsymbol{\alpha}$, $\widehat{\omega}_k = \sqrt{\mathrm{diag}(\Sigma_k)}$ and therefore, (S.2) becomes

$$V(\boldsymbol{x}) = \sum_{k=1}^{D} c \frac{\exp\{\tilde{\omega}_k^2/2 - a_k\}}{x_k} \Psi\left(\boldsymbol{a}_{-k} - a_k \mathbf{1}_{D-1} + \log(\boldsymbol{x}_{-k}/x_k); \boldsymbol{\mu}_k, \Sigma_k, \widehat{\boldsymbol{\alpha}}_k, \tau_k\right). \quad (S.3)$$

The normalizing constants are derived by solving $a_k = \log\left(\mathbb{E}\left[\exp\{Y_k\}\right]\right)$ for $k = 1, \ldots, D$. The moment generating function of $Y \sim \mathrm{ESN}(\mathbf{0}, \Sigma, \boldsymbol{\alpha}, 0)$ is given by

$$M(\boldsymbol{t}) = 2 \exp\left\{\tfrac{1}{2} \boldsymbol{t}^\top \Sigma \boldsymbol{t}\right\} \Phi\left(\boldsymbol{\delta}^\top \omega \boldsymbol{t}\right),$$

where $\boldsymbol{\delta}$ and $\omega$ are as defined above and considering $\boldsymbol{t} = \boldsymbol{e}_k$ yields

$$a_k = \log(2) + \tfrac{1}{2} \tilde{\omega}_k^2 + \log\left(\Phi\left(\tilde{\omega}_k \delta_k\right)\right).$$

Noting that $\tau_k = \tilde{\omega}_k \delta_k$ and

$$\boldsymbol{a}_{-k} - a_k \mathbf{1}_{D-1} + \log\left(\frac{\boldsymbol{x}_{-k}}{x_k}\right) = \log\left(\frac{\boldsymbol{x}_{-k}^{\circ\circ}}{x_k^{\circ\circ}}\right) + \frac{\tilde{\omega}_{-k}^2 - \tilde{\omega}_k^2 \mathbf{1}_{D-1}}{2},$$

where $\boldsymbol{x}^{\circ\circ} = (x_1^{\circ\circ}, \ldots, x_D^{\circ\circ})$, $x_k^{\circ\circ} = 2x_k \Phi(\tau_k; 0, 1)$, $k = 1, \ldots, D$, we can therefore re-write (S.3) as

$$V(\boldsymbol{x}) = \sum_{k=1}^{D} \frac{1}{x_k} \Psi\left(\log\left(\frac{\boldsymbol{x}_{-k}^{\circ\circ}}{x_k^{\circ\circ}}\right) + \frac{\tilde{\omega}_{-k}^2 - \tilde{\omega}_k^2 \mathbf{1}_{D-1}}{2}; \boldsymbol{\mu}_k, \Sigma_k, \widehat{\boldsymbol{\alpha}}_k, \tau_k\right)$$

$$= \sum_{k=1}^{D} \frac{1}{x_k} \frac{\Phi\left(\begin{bmatrix} \widehat{\omega}_k^{-1}\left(\log\left(\frac{\boldsymbol{x}_{-k}^{\circ\circ}}{x_k^{\circ\circ}}\right) + \frac{\tilde{\omega}_{-k}^2 - \tilde{\omega}_k^2 \mathbf{1}_{D-1}}{2} - \boldsymbol{\mu}_k\right) \\ \tau_k \end{bmatrix}; \begin{bmatrix} \mathbf{0} \\ 0 \end{bmatrix}, \begin{bmatrix} \overline{\Sigma}_k & -\frac{\overline{\Sigma}_k \widehat{\boldsymbol{\alpha}}_k}{(1+\widehat{\boldsymbol{\alpha}}_k^\top \overline{\Sigma}_k \widehat{\boldsymbol{\alpha}}_k)^{1/2}} \\ -\frac{\widehat{\boldsymbol{\alpha}}_k^\top \overline{\Sigma}_k}{(1+\widehat{\boldsymbol{\alpha}}_k^\top \overline{\Sigma}_k \widehat{\boldsymbol{\alpha}}_k)^{1/2}} & 1 \end{bmatrix}\right)}{\Phi(\tau_k; 0, 1)}$$

$$= \sum_{k=1}^{D} \frac{1}{x_k} \frac{\Phi\left(\begin{bmatrix} \widehat{\omega}_k^{-1}\left(\log\left(\frac{\boldsymbol{x}_{-k}^{\circ\circ}}{x_k^{\circ\circ}}\right) + \frac{\tilde{\omega}_{-k}^2 - \tilde{\omega}_k^2 \mathbf{1}_{D-1}}{2} - \boldsymbol{\mu}_k\right) \\ \tau_k \end{bmatrix}; \begin{bmatrix} \mathbf{0} \\ 0 \end{bmatrix}, \begin{bmatrix} \overline{\Sigma}_k & -\widehat{\omega}_k^{-1} A_k \omega \boldsymbol{\delta} \\ -\boldsymbol{\delta}^\top \omega A_k^\top \widehat{\omega}_k^{-1} & 1 \end{bmatrix}\right)}{\Phi(\tau_k; 0, 1)}.$$

which completes the proof of Theorem 2.

Notice that the expression above can be much simplified by re-parametrising the slant parameter $\boldsymbol{\alpha}$ by $\boldsymbol{\eta} = \omega^{-1}\boldsymbol{\alpha}$ which implies that $\omega\boldsymbol{\delta} = \Sigma\boldsymbol{\eta}\left(\sqrt{1+\boldsymbol{\eta}^\top\Sigma\boldsymbol{\eta}}\right)^{-1/2} \equiv \boldsymbol{\xi}$. As such, we have $\tau_k = \xi_k$ and $\widehat{\boldsymbol{\alpha}}_k = (1 - \boldsymbol{\xi}^\top A_k^\top \Sigma_k^{-1} A_k \boldsymbol{\xi})^{-1/2} \widehat{\omega}_k \Sigma_k^{-1} A_k \boldsymbol{\xi}$, and the exponent function can be written as

$$V(\boldsymbol{x}) = \sum_{k=1}^{D} \frac{1}{x_k} \frac{\Phi\left(\begin{bmatrix} \widehat{\omega}_k^{-1}\left(\log\left(\frac{\boldsymbol{x}_{-k}^{\circ\circ}}{x_k^{\circ\circ}}\right) + \frac{\tilde{\omega}_{-k}^2 - \tilde{\omega}_k^2 \mathbf{1}_{D-1}}{2} - \boldsymbol{\mu}_k\right) \\ \xi_k \end{bmatrix}; \begin{bmatrix} \mathbf{0} \\ 0 \end{bmatrix}, \begin{bmatrix} \overline{\Sigma}_k & -\widehat{\omega}_k^{-1} A_k \boldsymbol{\xi} \\ -\boldsymbol{\xi}^\top A_k^\top \widehat{\omega}_k^{-1} & 1 \end{bmatrix}\right)}{\Phi(\xi_k; 0, 1)}$$

$$= \sum_{k=1}^{D} \frac{1}{x_k} \frac{\Phi\left(\begin{bmatrix} \log\left(\frac{\boldsymbol{x}_{-k}^{\circ\circ}}{x_k^{\circ\circ}}\right) + \frac{\tilde{\omega}_{-k}^2 - \tilde{\omega}_k^2 \mathbf{1}_{D-1}}{2} - \boldsymbol{\mu}_k \\ \xi_k \end{bmatrix}; \begin{bmatrix} \mathbf{0} \\ 0 \end{bmatrix}, \begin{bmatrix} \Sigma_k & -A_k \boldsymbol{\xi} \\ -\boldsymbol{\xi}^\top A_k^\top & 1 \end{bmatrix}\right)}{\Phi(\xi_k; 0, 1)}.$$

## C.2 Proof of Proposition 1

The conditional probability (13) is written as

$$\Pr(\boldsymbol{W}_{\overline{B}_k} = \mathbf{0}_{D-k} \mid \boldsymbol{W}_{B_k} = r\boldsymbol{x}_{B_k}) = \Pr(\boldsymbol{Y}_{\overline{B}_k} = -\boldsymbol{\infty}_{D-k} \mid \boldsymbol{Y}_{B_k} = \log(r\boldsymbol{x}_{B_k}) + \boldsymbol{a}_{B_k}),$$

and since $\boldsymbol{Y}_{\overline{B}_k} \mid \boldsymbol{Y}_{B_k}$ is distributed according to the extended skew-Normal distribution, it directly implies that the above probability is zero and therefore the skewed Brown-Resnick process has no mass on $\partial\Omega$. It now remains to derive its density on $\Omega^\circ$.

Take $w = (w_1, \ldots, w_D) \in \mathbb{R}_+^D$ and let $w_k^{\circ\circ} = 2w_k \Phi(\tau_k; 0, 1)$, the density of $W$ is given by

$$f_W(w) = f_{\tilde{Y}}\left(\tfrac{\tilde{w}^2}{2} + \log(w^{\circ\circ})\right) \times \prod_{k=1}^{D} \tfrac{1}{w_k}$$

$$= \psi\left(\tfrac{\tilde{\omega}^2}{2} + \log(w^{\circ\circ}); 0, \Sigma, \alpha, 0\right) \times \prod_{k=1}^{D} \tfrac{1}{w_k}. \tag{S.4}$$

In order to complete the proof we need the following lemma.

**Lemma 1.** *The intensity function for the skewed Brown-Resnick process is*

$$\kappa(x) = \tfrac{2\Phi(\tilde{\tau};0,1)|\Sigma|^{-1/2}(\mathbf{1}^\top q)^{-1/2}}{(2\pi)^{(D-1)/2} \prod_{k=1}^{D} x_k} \exp\left\{-\tfrac{1}{2}\left[\log x^{\circ\circ\top} \mathcal{M} \log x^{\circ\circ} + \log x^{\circ\circ\top}\left(\tfrac{2q}{\mathbf{1}^\top q} + \mathcal{M}\tilde{\omega}^2\right) + \tfrac{q^\top \tilde{\omega}^2 - 1}{\mathbf{1}^\top q} + \tfrac{1}{4}\tilde{\omega}^{2,\top}\mathcal{M}\tilde{\omega}^2\right]\right\},$$

*where $q = \Sigma^{-1}\mathbf{1}$, $\mathcal{M} = \Sigma^{-1} - qq^\top/\mathbf{1}^\top q$ and*

$$\tilde{\tau} = \left(1 + \tfrac{(\alpha^\top \omega^{-1}\mathbf{1})^2}{\mathbf{1}^\top q}\right)^{-1/2} \alpha^\top \omega^{-1}\left[\left(\mathbb{I} - \tfrac{\mathbf{1}q^\top}{\mathbf{1}^\top q}\right)\left(\log x^{\circ\circ} + \tfrac{\tilde{\omega}^2}{2}\right) + \tfrac{\mathbf{1}}{\mathbf{1}^\top q}\right].$$

*Proof.* By plugging (S.4) into (10), we get

$$\kappa(x) = \int_0^\infty r^D \psi\left(\tfrac{\tilde{\omega}^2}{2} + \log(x^{\circ\circ}) + \log(r); 0, \Sigma, \alpha, 0\right) \times \prod_{k=1}^{D} \tfrac{1}{rx_k} dr$$

$$= \int_0^\infty \psi\left(\tfrac{\tilde{\omega}^2}{2} + \log(x^{\circ\circ}) + \log(r); 0, \Sigma, \alpha, 0\right) \times \prod_{k=1}^{D} \tfrac{1}{x_k} dr$$

$$= \int_{\mathbb{R}} e^u \psi\left(\tfrac{\tilde{\omega}^2}{2} + \log(x^{\circ\circ}) + u; 0, \Sigma, \alpha, 0\right) \times \prod_{k=1}^{D} \tfrac{1}{x_k} du$$

$$= \int_{\mathbb{R}} \tfrac{e^u 2}{\prod_{k=1}^{D} x_k} \phi\left(\tfrac{\tilde{\omega}^2}{2} + \log(x^{\circ\circ}) + u; 0, \Sigma\right) \Phi\left(\alpha^\top \omega^{-1}\left(\tfrac{\tilde{\omega}^2}{2} + \log(x^{\circ\circ}) + u\right); 0, 1\right) du$$

Following [Wadsworth and Tawn (2014)](#), we have

$$e^u \phi\left(\tfrac{\tilde{\omega}^2}{2} + \log(x^{\circ\circ}) + u; 0, \Sigma\right) = \phi\left(u; \mu_u, \omega_u^2\right) \tfrac{|\Sigma|^{-1/2}(\mathbf{1}^\top q)^{-1/2}}{(2\pi)^{(D-1)/2}}$$

$$\times \exp\left\{-\tfrac{1}{2}\left[\log x^{\circ\circ\top}\mathcal{M}\log x^{\circ\circ} + \log x^{\circ\circ\top}\left(\tfrac{2q}{\mathbf{1}^\top q} + \mathcal{M}\tilde{\omega}^2\right)\right]\right\}$$

$$\times \exp\left\{-\tfrac{1}{2}\left(\tfrac{q^\top \tilde{\omega}^2 - 1}{\mathbf{1}^\top q} + \tfrac{1}{4}\tilde{\omega}^{2,\top}\mathcal{M}\tilde{\omega}^2\right)\right\}$$

where $q = \Sigma^{-1}\mathbf{1}$, $\mu_u = -\frac{\log x^{\circ\circ} q + \frac{1}{2}(\tilde{\omega}^2)^\top q - 1}{\mathbf{1}^\top q}$, $\omega_u = (\mathbf{1}^\top q)^{-1/2}$, and $\mathcal{M} = \Sigma^{-1} - qq^\top/\mathbf{1}^\top q$. In addition, note that

$$\Phi\left(\alpha^\top \omega^{-1}\left(\frac{\tilde{\omega}^2}{2} + \log(x^{\circ\circ}) + u\right); 0, 1\right) = \Phi\left(\alpha_u\left(\frac{u - \mu_u}{\omega_u}\right) + \alpha_{0,u}; 0, 1\right) \frac{\Phi(\tilde{\tau}; 0, 1)}{\Phi(\tilde{\tau}; 0, 1)}$$

where $\alpha_u = \omega_u \alpha^\top \omega^{-1} \mathbf{1}$, $\alpha_{0,u} = \alpha^\top \omega^{-1}\left(I - \omega_u^2 \mathbf{1}\mathbf{1}^\top \Sigma^{-1}\right)\left(\log x^{\circ\circ} + \tilde{\omega}^2/2\right) - \omega_u^2 \alpha^\top \omega^{-1}\mathbf{1}$ and

$$\tilde{\tau} = \alpha_{0,u}(1 + \alpha_u^2)^{-1/2} = \left(1 + \frac{(\alpha^\top \omega^{-1}\mathbf{1})^2}{\mathbf{1}^\top q}\right)^{-1/2} \alpha^\top \omega^{-1}\left[\left(\mathbb{I} - \frac{\mathbf{1}q^\top}{\mathbf{1}^\top q}\right)\left(\log x^{\circ\circ} + \frac{\tilde{\omega}^2}{2}\right) + \frac{\mathbf{1}}{\mathbf{1}^\top q}\right]. \quad (S.5)$$

Therefore, by matching the integrand with a univariate extended skew normal density function with respect to $u$, we have,

$$\begin{aligned}
\kappa(x) &= \int_{\mathbb{R}} \phi\left(u; \mu_u, \omega_u^2\right) \Phi\left(\alpha_u\left(\frac{u-\mu_u}{\omega_u}\right) + \alpha_{0,u}; 0, 1\right) \frac{\Phi(\tilde{\tau}; 0, 1)}{\Phi(\tilde{\tau}; 0, 1)} \frac{2|\Sigma|^{-1/2}(\mathbf{1}^\top q)^{-1/2}}{(2\pi)^{(D-1)/2} \prod_{k=1}^D x_k} \\
&\quad \times \exp\left\{-\frac{1}{2}\left[\log x^{\circ\circ\top} \mathcal{M} \log x^{\circ\circ} + \log x^{\circ\circ\top}\left(\frac{2q}{\mathbf{1}^\top q} + \mathcal{M}\tilde{\omega}^2\right)\right]\right\} \\
&\quad \times \exp\left\{-\frac{1}{2}\left(\frac{q^\top \tilde{\omega}^2 - 1}{\mathbf{1}^\top q} + \frac{1}{4}\tilde{\omega}^{2,\top} \mathcal{M}\tilde{\omega}^2\right)\right\} du \\
&= \int_{\mathbb{R}} \phi\left(u; \mu_u, \omega_u^2\right) \Phi\left(\alpha_u\left(\frac{u-\mu_u}{\omega_u}\right) + \alpha_{0,u}; 0, 1\right) \frac{1}{\Phi(\tilde{\tau}; 0, 1)} du \\
&\quad \times \frac{2\Phi(\tilde{\tau}; 0, 1)|\Sigma|^{-1/2}(\mathbf{1}^\top q)^{-1/2}}{(2\pi)^{(D-1)/2} \prod_{k=1}^D x_k} \\
&\quad \times \exp\left\{-\frac{1}{2}\left[\log x^{\circ\circ\top} \mathcal{M} \log x^{\circ\circ} + \log x^{\circ\circ\top}\left(\frac{2q}{\mathbf{1}^\top q} + \mathcal{M}\tilde{\omega}^2\right)\right]\right\} \\
&\quad \times \exp\left\{-\frac{1}{2}\left(\frac{q^\top \tilde{\omega}^2 - 1}{\mathbf{1}^\top q} + \frac{1}{4}\tilde{\omega}^{2,\top} \mathcal{M}\tilde{\omega}^2\right)\right\} \\
&= \frac{2\Phi(\tilde{\tau}; 0, 1)|\Sigma|^{-1/2}(\mathbf{1}^\top q)^{-1/2}}{(2\pi)^{(D-1)/2} \prod_{k=1}^D x_k} \exp\left\{-\frac{1}{2}\left[\log x^{\circ\circ\top} \mathcal{M} \log x^{\circ\circ} + \log x^{\circ\circ\top}\left(\frac{2q}{\mathbf{1}^\top q} + \mathcal{M}\tilde{\omega}^2\right) + \frac{q^\top \tilde{\omega}^2 - 1}{\mathbf{1}^\top q} + \frac{1}{4}\tilde{\omega}^{2,\top} \mathcal{M}\tilde{\omega}^2\right]\right\},
\end{aligned}$$

which gives the desired expression of the intensity function. $\square$

### C.3 Partial derivatives of the exponent function

**Lemma 2.** *The partial derivatives of the exponent function for the skewed Brown-Resnick process are given as*

$$-V_{B_k}(x) = -\frac{V_{B_k}(x_{B_k}, \infty_{D-k})}{\Phi\left(\tilde{\alpha}_0\left(1 + \tilde{\alpha}^\top \overline{\tilde{\Sigma}}\tilde{\alpha}\right)^{-1/2}; 0, 1\right)} \\
\times \Phi\left(\begin{bmatrix} \tilde{\omega}_{(\overline{B}_k)}^{-1}(\log x_{\overline{B}_k}^{\circ\circ} - \tilde{\mu}) \\ \tilde{\alpha}_0(1 + \tilde{\alpha}^\top \overline{\tilde{\Sigma}}\tilde{\alpha})^{-1/2} \end{bmatrix}; \begin{bmatrix} 0 \\ 0 \end{bmatrix} \begin{bmatrix} \overline{\tilde{\Sigma}} & -(1 + \tilde{\alpha}^\top \overline{\tilde{\Sigma}}\tilde{\alpha})^{-1/2}\overline{\tilde{\Sigma}}\tilde{\alpha} \\ -(1 + \tilde{\alpha}^\top \overline{\tilde{\Sigma}}\tilde{\alpha})^{-1/2}\overline{\tilde{\Sigma}}\tilde{\alpha} & 1 \end{bmatrix}\right),$$

where $-V_{B_k}(x_{B_k}, \infty_{D-k})$ is the marginal intensity function, which equals the intensity function in Lemma 1 with parameters replaced by the marginal covariance matrix $\Sigma_{B_k B_k}$ and the marginal slant parameter, $\alpha_{(B_k)}$, defined as,

$$\alpha_{(B_k)} = \left(1 + \alpha_{\overline{B}_k}^\top \left(\overline{\Sigma}_{\overline{B}_k \overline{B}_k} - \overline{\Sigma}_{\overline{B}_k B_k} \overline{\Sigma}_{B_k B_k}^{-1} \overline{\Sigma}_{B_k \overline{B}_k}\right) \alpha_{\overline{B}_k}\right)^{-1/2} \left(\alpha_{B_k} + \overline{\Sigma}_{B_k B_k}^{-1} \overline{\Sigma}_{B_k \overline{B}_k} \alpha_{\overline{B}_k}\right),$$

and $\tilde{\Sigma} = \left(\mathcal{M}_{\overline{B}_k \overline{B}_k}\right)^{-1}$, $\overline{\tilde{\Sigma}} = \omega_{(\overline{B}_k)}^{-1} \tilde{\Sigma} \omega_{(\overline{B}_k)}^{-1}$, $\omega_{(\overline{B}_k)} = \sqrt{\text{diag}(\tilde{\Sigma})}$,

$\tilde{\alpha} = \left(1 + \frac{(\alpha^\top \omega^{-1} \mathbf{1})^2}{\mathbf{1}^\top q}\right)^{-1/2} \omega_{(\overline{B}_k)} K_{01} \left(\mathbb{I} - \frac{\mathbf{1} q^\top}{\mathbf{1}^\top q}\right) \omega^{-1} \alpha,$

$\tilde{\alpha}_0 = \boldsymbol{\beta}^\top K_{10}^\top (\log x_{B_k}^{\circ\circ} + K_{10} \tilde{\omega}^2/2) + \boldsymbol{\beta}^\top K_{01}^\top (\tilde{\mu} + K_{01} \tilde{\omega}^2/2) + \left(1 + \frac{(\alpha^\top \omega^{-1} \mathbf{1})^2}{\mathbf{1}^\top q}\right)^{-1/2} \frac{\alpha^\top \omega^{-1} \mathbf{1}}{\mathbf{1}^\top q},$

$K_{10} = (\mathbb{I}_k, \mathbf{0}_{k, D-k}),\ K_{01} = (\mathbf{0}_{D-k, k}, \mathbb{I}_{D-k}),\ \boldsymbol{\beta} = \left(1 + \frac{(\alpha^\top \omega^{-1} \mathbf{1})^2}{\mathbf{1}^\top q}\right)^{-1/2} \left(\mathbb{I} - \frac{q \mathbf{1}^\top}{\mathbf{1}^\top q}\right) \omega^{-1} \alpha,$

$\tilde{\mu} = -\tilde{\Sigma} \left(\mathcal{M}_{\overline{B}_k B_k} \log x_{B_k}^{\circ\circ} + K_{01} q/(\mathbf{1}^\top q) + K_{01} \mathcal{A} \tilde{\omega}^2/2\right).$

*Proof.* We can write

$$-V_{B_k}(x) = -V_{B_k}(x_{B_k}, \infty_{D-k}) \int_0^{x_{\overline{B}_k}} \frac{V_{B_D}(x_{B_k}, y)}{V_{B_k}(x_{B_k}, \infty_{D-k})} dy,$$

which involves two parts, the marginal intensity function $-V_{B_k}(x_{B_k}, \infty_{D-k})$ and the integral of the conditional intensity function over $x_{\overline{B}_k}$. As the marginal distribution of the constructive spectral component $Y_{B_k}$ in Theorem 3 from the main manuscript follows the distribution $\text{ESN}(\mathbf{0}, \Sigma_{B_k B_k}, \alpha_{(B_k)}, 0)$ (Azzalini, 2013, Chapter 5.1.4), where $\alpha_{(B_k)}$ is defined above, the marginal intensity function $-V_{B_k}(x_{B_k}, \infty_{D-k})$ has the same expression as in Lemma 1 but with parameters, $\Sigma$ and $\alpha$, replaced by their marginal counterparts, $\Sigma_{B_k B_k}$ and $\alpha_{(B_k)}$.

The conditional intensity function $V_{B_D}(x)/V_{B_k}(x_{B_k}, \infty_{D-k})$ is indeed a valid density function (Wadsworth and Tawn, 2013) with respect to $x_{\overline{B}_k}$. Treating $x_{B_k}$ as constant, the conditional intensity function can be expressed as

$\frac{V_{B_D}(x)}{V_{B_k}(x_{B_k}, \infty_{D-k})}$

$\propto \kappa(x)$

$= \frac{2\Phi(\tilde{\tau}; 0, 1) |\Sigma|^{-1/2} (\mathbf{1}^\top q)^{-1/2}}{(2\pi)^{(D-1)/2} \prod_{k=1}^D x_k} \exp\left\{-\frac{1}{2} \left[\log x^{\circ\circ\top} \mathcal{M} \log x^{\circ\circ} + \log x^{\circ\circ\top} \left(\frac{2q}{\mathbf{1}^\top q} + \mathcal{M} \tilde{\omega}^2\right) + \frac{q^\top \tilde{\omega}^2 - 1}{\mathbf{1}^\top q} + \frac{1}{4} \tilde{\omega}^{2,\top} \mathcal{M} \tilde{\omega}^2\right]\right\}.$

By comparing above equation with the multivariate extended skew normal density function, we can show that the conditional intensity function with respect to $\log x_{\overline{B}_k}^{\circ\circ}$ is a density function for a multivariate extended skew normal random vector, $\log X_{\overline{B}_k}^{\circ\circ}$, with scale matrix $\tilde{\Sigma} = \left(\mathcal{M}_{\overline{B}_k \overline{B}_k}\right)^{-1}$, location parameter $\tilde{\mu} = -\tilde{\Sigma}\left(\mathcal{M}_{\overline{B}_k B_k} \log(x_{B_k}^{\circ\circ}) + K_{01} q/(\mathbf{1}^\top q) + K_{01} \mathcal{M} \tilde{\omega}^2/2\right)$, where $K_{01} = (\mathbf{0}_{D-k,k}, \mathbb{I}_{D-k})$, slant parameter $\tilde{\alpha} = \left(1 + \frac{(\alpha^\top \omega^{-1} \mathbf{1})^2}{\mathbf{1}^\top q}\right)^{-1/2} \omega_{(\overline{B}_k)} K_{01}\left(\mathbb{I} - \frac{\mathbf{1} q^\top}{\mathbf{1}^\top q}\right) \omega^{-1} \alpha$, where $\omega_{(\overline{B}_k)} = \sqrt{\text{diag}(\tilde{\Sigma})}$, and parameter

$$\tilde{\alpha}_0 = \beta^\top K_{10}^\top (\log x_{B_k}^{\circ\circ} + K_{10}\tilde{\omega}^2/2) + \beta^\top K_{01}^\top (\tilde{\mu} + K_{01}\tilde{\omega}^2/2) + \left(1 + \frac{(\alpha^\top \omega^{-1}\mathbf{1})^2}{\mathbf{1}^\top q}\right)^{-1/2} \frac{\alpha^\top \omega^{-1}\mathbf{1}}{\mathbf{1}^\top q},$$

where $K_{10} = (\mathbb{I}_k, \mathbf{0}_{k,D-k})$ and $\beta = \left(1 + \frac{(\alpha^\top \omega^{-1}\mathbf{1})^2}{\mathbf{1}^\top q}\right)^{-1/2} \left(\mathbb{I} - \frac{q\mathbf{1}^\top}{\mathbf{1}^\top q}\right) \omega^{-1} \alpha$. Therefore, we have

$$\frac{V_{B_D}(\boldsymbol{x})}{V_{B_k}(\boldsymbol{x}_{B_k}, \infty_{D-k})} = \frac{|\tilde{\Sigma}|^{-1/2}(2\pi)^{-(D-k)/2}}{\Phi\left(\tilde{\alpha}_0\left(1+\tilde{\alpha}^\top \omega_{(\overline{B}_k)}^{-1}\tilde{\Sigma}\omega_{(\overline{B}_k)}^{-1}\tilde{\alpha}\right)^{-1/2}; 0, 1\right) \prod_{i \in \overline{B}_k} x_i^{\circ\circ}} \Phi\left(\tilde{\alpha}_0 + \tilde{\alpha}^\top \omega_{(\overline{B}_k)}^{-1}\left(\log x_{\overline{B}_k}^{\circ\circ} - \tilde{\mu}\right); 0, 1\right)$$

$$\times \exp\left\{-\tfrac{1}{2}(\log x_{\overline{B}_k}^{\circ\circ} - \tilde{\mu})^\top \mathcal{M}_{\overline{B}_k \overline{B}_k} (\log x_{\overline{B}_k}^{\circ\circ} - \tilde{\mu})\right\}.$$

Then, by using the cumulative distribution function of $\log X_{\overline{B}_k}^{\circ\circ}$, the integral can be computed as

$$\int_0^{x_{\overline{B}_k}} \frac{V_{B_D}(\boldsymbol{x}_{B_k}, \boldsymbol{y})}{V_{B_k}(\boldsymbol{x}_{B_k}, \infty_{D-k})} d\boldsymbol{y} = \Pr[X_{\overline{B}_k} \leq x_{\overline{B}_k}]$$

$$= \Pr[\log X_{\overline{B}_k} \leq \log x_{\overline{B}_k}]$$

$$= \Pr[\log X_{\overline{B}_k}^{\circ\circ} \leq \log x_{\overline{B}_k}^{\circ\circ}]$$

$$= \frac{\Phi\left(\begin{bmatrix} \tilde{\omega}_{(\overline{B}_k)}^{-1}(\log x_{\overline{B}_k}^{\circ\circ} - \tilde{\mu}) \\ \tilde{\alpha}_0(1+\tilde{\alpha}^\top \overline{\tilde{\Sigma}}\tilde{\alpha})^{-1/2} \end{bmatrix}; \begin{bmatrix} \mathbf{0} \\ 0 \end{bmatrix}, \begin{bmatrix} \overline{\tilde{\Sigma}} & -(1+\tilde{\alpha}^\top \overline{\tilde{\Sigma}}\tilde{\alpha})^{-1/2}\overline{\tilde{\Sigma}}\tilde{\alpha} \\ -(1+\tilde{\alpha}^\top \overline{\tilde{\Sigma}}\tilde{\alpha})^{-1/2}\overline{\tilde{\Sigma}}\tilde{\alpha} & 1 \end{bmatrix}\right)}{\Phi(\tilde{\alpha}_0(1+\tilde{\alpha}^\top \overline{\tilde{\Sigma}}\tilde{\alpha})^{-1/2})}$$

where $\overline{\tilde{\Sigma}} = \omega_{(\overline{B}_k)}^{-1} \tilde{\Sigma} \omega_{(\overline{B}_k)}^{-1}$. Plugging the above results, the partial derivative of the exponent function are given by

$$-V_{B_k}(\boldsymbol{x}) = \frac{-V_{B_k}(\boldsymbol{x}_{B_k}, \infty_{D-k})}{\Phi(\tilde{\alpha}_0(1+\tilde{\alpha}^\top \overline{\tilde{\Sigma}}\tilde{\alpha})^{-1/2})}$$

$$\times \Phi\left(\begin{bmatrix} \tilde{\omega}_{(\overline{B}_k)}^{-1}(\log x_{\overline{B}_k}^{\circ\circ} - \tilde{\mu}) \\ \tilde{\alpha}_0(1+\tilde{\alpha}^\top \overline{\tilde{\Sigma}}\tilde{\alpha})^{-1/2} \end{bmatrix}; \begin{bmatrix} \mathbf{0} \\ 0 \end{bmatrix}, \begin{bmatrix} \overline{\tilde{\Sigma}} & -(1+\tilde{\alpha}^\top \overline{\tilde{\Sigma}}\tilde{\alpha})^{-1/2}\overline{\tilde{\Sigma}}\tilde{\alpha} \\ -\left(1+\tilde{\alpha}^\top \overline{\tilde{\Sigma}}\tilde{\alpha}\right)^{-1/2}\overline{\tilde{\Sigma}}\tilde{\alpha} & 1 \end{bmatrix}\right)$$

which concludes the proof. □

From Lemma 2, the limit of the partial derivative $-V_{B_k}(x)$ for any index set $B_k \neq \emptyset$ and $B_k \subsetneq \{1, \ldots, D\}$ as $x_{\overline{B}_k} \to 0$ is indeed zero.

## C.4 Exact simulation algorithm

The exact simulation algorithm for skewed Brown-Resnick process arises from Algorithm 2 in Dombry *et al.* (2016), which requires simulating a process $W/W(s_0)$ with distribution $P_{s_0}(A) = E_W[\mathbb{1}_{\{W/W(s_0) \in A\}} W(s_0)]$ and where $W = (W_0, W_1, \ldots, W_D) \equiv (W(s_0), W(s_1), \ldots, W(s_D))$ is defined as in Theorem 2 to the exception of being $D+1$ dimensional. The distribution $P_{s_0}$ defines the joint distribution of $W_i/W_0, i = 1, \ldots, D$ under the transformed probability measure $\widehat{\Pr} = W(s_0)\mathrm{d}\Pr$. The moment generating function of $Y_i \equiv Y(s_i), i = 0, \ldots, D$ evaluated at $t = (t_0, t_1, \ldots, t_D)$ can be calculated under the transformed probability measure $\widehat{\Pr}$ as

$$E\left[\exp\left\{Y_0 - a_0 + \sum_{i=0}^{D} t_i Y_i\right\}\right] = 2\exp\left\{\tfrac{1}{2}(1+t_0, t_{B_D})^\top \Sigma (1+t_0, t_{B_D}) - a_0\right\} \Phi(\delta^\top \omega (1+t_0, t_{B_D}))$$
$$= 2\exp\{\tfrac{1}{2}(\Sigma_{0;0} + t^\top \Sigma t) + \Sigma_{0;\cdot} t - a_0\} \Phi\left(\delta_0 \omega_{0;0} + \delta^\top \omega t\right)$$

where the matrices $\omega$ and $\Sigma$ can be decomposed as $\Sigma = \begin{bmatrix} \Sigma_{00} & \Sigma_{B_D;0} \\ \Sigma_{B_D;0}^\top & \Sigma_{B_D;B_D} \end{bmatrix}$. Plugging in the expression for $a_0$ given in Theorem 2 leads to

$$E\left[\exp\left\{Y_0 - a_0 + \sum_{i=0}^{D} t_i Y_i\right\}\right] = \exp\{\tfrac{1}{2} t^\top \Sigma t + \Sigma_{\cdot;0} t^\top\} \frac{\Phi(\delta_0 \omega_{00} + \delta^\top \omega t)}{\Phi(\delta_0 \omega_{00})},$$

which corresponds to the moment generating function of the extended skew-normal random vector $(Y^*(s_0), \ldots, Y^*(s_D))$ with location parameter $\Sigma_{\cdot;0}$, scale parameter $\Sigma$, slant parameter $\alpha$, and the extended parameter $\tau_0 = \delta_0 \omega_{00}$. The distribution of $P_{s_0}$ therefore becomes

$$\exp\{Y^*(s_i) - Y^*(s_0) - a_i + a_0\}, i = 0, \ldots, D.$$

Details about random generation from the multivariate extended skew-normal can be found in Azzalini (2013, Chapter 5.3.3). The pseudo-code of the exact simulation algorithms is summarised in Algorithm 1.

**Algorithm 1:** Exact simulation from the skewed Brown-Resnick process at location $s_i, i = 1, \ldots, D$.

**Input:** slant parameter $\boldsymbol{\alpha} \in \mathbb{R}^D$, $D \times D$ scale matrix $\Sigma$
**Output:** $Z(s_j), j = 1, \ldots, D$.

1. Compute $\boldsymbol{a}$ and $\boldsymbol{\delta}$ as defined in Theorem 2.
2. Simulate $E_0 \sim \text{Exp}(1)$ and set $z = 1/E_0$.
3. Simulate $Y^*(s)$ from an extended skew normal with location parameter $\Sigma_{\cdot;1}$, scale matrix $\Sigma$, slant parameter $\boldsymbol{\alpha}$ and the extended parameter $\tau_1 = \delta_1 \omega_{11}$.
4. Set $W(s) = \exp(Y(s) - Y(s_1) + a_1 - \boldsymbol{a})$
5. Set $Z(s_j) = W(s_j) * z$ for $j = 1, \ldots, D$.
6. **for** $i = 2, \ldots, D$ **do**
7.     Simulate $E_0 \sim \text{Exp}(1)$ and set $z = 1/E_0$.
8.     **while** $z > Z(s_i)$ **do**
9.         Simulate $Y(s)$ from an extended skew normal with location parameter $\Sigma_{\cdot;i}$, scale matrix $\Sigma$, slant parameter $\boldsymbol{\alpha}$ and extended parameter $\tau_i = \delta_i \omega_{ii}$
10.        Set $W(s) = \exp(Y(s) - Y(s_i) + a_i - \boldsymbol{a})$
11.        **if** $W(s_j) \times z < Z(s_j), \forall j = 1, \ldots, i-1$ **then**
12.           Set $Z(s_j) = \max(Z(s_j), W(s_j) \times z), j = 1, \ldots, D$.
13.        Simulate $E_1 \sim \text{Exp}(1)$, Set $E_0 = E_0 + E_1$ and $z = 1/E_0$.

# D   Truncated extremal-t processes

## D.1   Proof of Theorem 3

Let $W = \tilde{Y}^\nu / \boldsymbol{a}$, where $\tilde{Y}$ is the random vector associated with the process $\tilde{Y}(s)$, i.e., $\tilde{Y} \sim \mathcal{N}_{[\mathbf{0},\infty]}(\mathbf{0}, \Sigma)$ with unit variance and therefore $\Sigma$ represent the correlation matrix, and $\boldsymbol{a} = (a_1, \ldots, a_D)$, $a_k = \mathbb{E}[\tilde{Y}_k^\nu]$, $k = 1, \ldots, D$, and $\nu > 0$.

By applying the first half of (2), we have

$$V(\boldsymbol{x}) = \sum_{k=1}^{D} \frac{1}{a_k x_k} \int_0^\infty u_k^\nu \int_{\boldsymbol{0}}^{u_k \left(\frac{\boldsymbol{x}_{-k}^\circ}{x_k^\circ}\right)^{1/\nu}} \frac{\phi(\boldsymbol{u};\boldsymbol{0},\Sigma)}{\Phi^\circ(\infty;\boldsymbol{0},\Sigma)} \mathrm{d}\boldsymbol{u}_{-k} \mathrm{d}u_k$$

$$= \sum_{k=1}^{D} \frac{1}{a_k x_k \Phi^\circ(\infty;\boldsymbol{0},\Sigma)} \int_0^\infty u_k^\nu \phi(u_k;0,1) \int_{\boldsymbol{0}}^{u_k \left(\frac{\boldsymbol{x}_{-k}^\circ}{x_k^\circ}\right)^{1/\nu}} \phi(\boldsymbol{u}_{-k};\Sigma_{-k,k} u_k, \mathbb{I}_{D-1} - \Sigma_{-k,k}\Sigma_{k,-k}) \mathrm{d}\boldsymbol{u}_{-k} \mathrm{d}u_k$$

$$= \sum_{k=1}^{D} \frac{1}{a_k x_k \Phi^\circ(\infty;\boldsymbol{0},\Sigma)} \int_0^\infty u_k^\nu \phi(u_k;0,1) \Phi^\circ\left(u_k \left(\frac{\boldsymbol{x}_{-k}^\circ}{x_k^\circ}\right)^{1/\nu}; \Sigma_{-k,k} u_k, \mathbb{I}_{D-1} - \Sigma_{-k,k}\Sigma_{k,-k}\right) \mathrm{d}u_k$$

$$= \sum_{k=1}^{D} \frac{2^{(\nu-2)/2} \Gamma((\nu+1)/2)}{\Phi^\circ(\infty;\boldsymbol{0},\Sigma) a_j x_j \sqrt{\pi}} T_{\nu+1}^\circ\left(\left(\frac{\boldsymbol{x}_{-k}^\circ}{x_k^\circ}\right)^{1/\nu}; \Sigma_{-k,k}, \frac{\mathbb{I}_{D-1} - \Sigma_{-k,k}\Sigma_{k,-k}}{\nu+1}\right), \tag{S.6}$$

where $\boldsymbol{x}^\circ = \boldsymbol{x}\boldsymbol{a}$, $\mathbb{I}_{D-1}$ is the $(D-1)$-dimensional identity matrix, and the index $-k$ represents the set $B_D \setminus \{k\}$.

The normalising constants $a_k$ can be either derived from the fact that $V(\infty, \ldots, \infty, x_k, \infty, \ldots, \infty) = 1/x_k$ or by comparing the integral above to the following integral

$$a_k = E[\tilde{Y}_k^\nu] = \int_0^\infty u_k^\nu \int_{\boldsymbol{0}}^\infty \frac{\phi(\boldsymbol{u};\boldsymbol{0},\Sigma)}{\Phi^\circ(\infty;\boldsymbol{0},\Sigma)} \mathrm{d}\boldsymbol{u}_{-k} \mathrm{d}u_k.$$

The use of either way will lead us to

$$a_k = \frac{2^{(\nu-2)/2} \Gamma((\nu+1)/2)}{\Phi^\circ(\infty;\boldsymbol{0},\Sigma) \sqrt{\pi}} T_{\nu+1}^\circ\left(\infty_{D-1}; \Sigma_{-k,k}, \frac{\mathbb{I}_{D-1} - \Sigma_{-k,k}\Sigma_{k,-k}}{\nu+1}\right).$$

Plugging the expression for $a_k$ in (S.6), we have

$$V(\boldsymbol{x}) = \sum_{k=1}^{D} \frac{T_{\nu+1}^\circ\left(\left(\frac{\boldsymbol{x}_{-k}^\circ}{x_k^\circ}\right)^{1/\nu}; \Sigma_{-k,k}, \frac{\mathbb{I}_{D-1} - \Sigma_{-k,k}\Sigma_{k,-k}}{\nu+1}\right)}{x_k T_{\nu+1}^\circ\left(\infty_{D-1}; \Sigma_{-k,k}, \frac{\mathbb{I}_{D-1} - \Sigma_{-k,k}\Sigma_{k,-k}}{\nu+1}\right)}.$$

## D.2 Proof of Proposition 2

From the distribution of $\tilde{\boldsymbol{Y}}$ and therefore $\boldsymbol{W}$, the conditional probability in (13) is written as

$$\Pr(\boldsymbol{W}_{\overline{B}_k} = \boldsymbol{0}_{D-k} \mid \boldsymbol{W}_{B_k} = r\boldsymbol{x}_{B_k}) = \Phi^\circ\left(\boldsymbol{0}_{D-k}; \Sigma_{\overline{B}_k} \Sigma_{B_k B_k}^{-1} \left(r\boldsymbol{x}_{B_k}^\circ\right)^{1/\nu}, \Sigma_{\overline{B}_k \overline{B}_k} - \Sigma_{\overline{B}_k B_k} \Sigma_{B_k B_k}^{-1} \Sigma_{\overline{B}_k B_k}\right)$$

$$= 0,$$

which, by Theorem 1, implies that the truncated extramal-t process has no mass on $\partial\Omega$. It now remains to derive its density on $\Omega^\circ$.

Take $\boldsymbol{w} = (w_1, \ldots, w_D) \in \mathbb{R}_+^D$ and let $\boldsymbol{w}^\circ = \boldsymbol{w}\boldsymbol{a}$, the density of $\boldsymbol{W}$ is given by

$$
\begin{aligned}
f_W(\boldsymbol{w}) &= f_{\tilde{Y}}\left((\boldsymbol{w}^\circ)^{1/\nu}\right) \times \prod_{k=1}^{D} \left\{\tfrac{a_k}{\nu}\left(w_k^\circ\right)^{1/\nu-1}\right\} \\
&= \frac{\phi\left((\boldsymbol{w}^\circ)^{1/\nu}; \boldsymbol{0}, \Sigma\right)}{\Phi^\circ(\infty; \boldsymbol{0}, \Sigma)} \times \prod_{k=1}^{D} \left\{\tfrac{a_k}{\nu}\left(w_k^\circ\right)^{1/\nu-1}\right\}.
\end{aligned} \tag{S.7}
$$

In order to complete the proof we need the following lemma.

**Lemma 3.** *The intensity function for the truncated extremal-t process is*

$$
\kappa(\boldsymbol{x}) = \frac{\prod_{k=1}^{D}\left(a_k x_k^{1-\nu}\right)^{1/\nu}}{\Phi^\circ(\infty; \boldsymbol{0}, \Sigma)\nu^{D-1}|\Sigma|^{1/2}\pi^{D/2}2^{(2-\nu)/2}} \left\{\left[(\boldsymbol{x}^\circ)^{1/\nu}\right]^\top \Sigma^{-1} \left[(\boldsymbol{x}^\circ)^{1/\nu}\right]\right\}^{-(\nu+D)/2} \Gamma\left(\tfrac{\nu+D}{2}\right).
$$

*Proof.* By plugging (S.7) into (10), we get

$$
\begin{aligned}
\kappa(\boldsymbol{x}) &= \int_0^\infty r^D \frac{\phi((r\boldsymbol{x}^\circ)^{1/\nu}; \boldsymbol{0}, \Sigma)}{\Phi^\circ(\infty; \boldsymbol{0}, \Sigma)} \prod_{k=1}^{D} \left\{\tfrac{a_k}{\nu}(rx_k^\circ)^{1/\nu-1}\right\} dr \\
&= \frac{\prod_{k=1}^{D}\left\{a_k^{1/\nu} x_k^{1/\nu-1}\right\}}{\Phi^\circ(\infty; \boldsymbol{0}, \Sigma)\nu^D} \int_0^\infty r^{D/\nu} \phi((r\boldsymbol{x}^\circ)^{1/\nu}; \boldsymbol{0}, \Sigma) dr \\
&= \frac{\prod_{k=1}^{D}\left(a_k x_k^{\nu-1}\right)^{1/\nu}}{\Phi^\circ(\infty; \boldsymbol{0}, \Sigma)\nu^D|\Sigma|^{1/2}\pi^{D/2}2^{(2-\nu)/2}} \int_0^\infty r^{D/\nu} \exp\left\{-\tfrac{r^{2/\nu}}{2}\left[(\boldsymbol{x}^\circ)^{1/\nu}\right]^\top \Sigma^{-1}\left[(\boldsymbol{x}^\circ)^{1/\nu}\right]\right\} dr \\
&= \frac{\prod_{k=1}^{D}\left(a_k x_k^{1-\nu}\right)^{1/\nu}}{\Phi^\circ(\infty; \boldsymbol{0}, \Sigma)\nu^{D-1}|\Sigma|^{1/2}\pi^{D/2}2^{(2-\nu)/2}} \left\{\left[(\boldsymbol{x}^\circ)^{1/\nu}\right]^\top \Sigma^{-1}\left[(\boldsymbol{x}^\circ)^{1/\nu}\right]\right\}^{-(\nu+D)/2} \Gamma\left(\tfrac{\nu+D}{2}\right),
\end{aligned}
$$

which completes the proof. □

By Lemma 3, the density on the interior of the simplex, $\Omega^\circ$ is therefore given by

$$
h(\boldsymbol{w}) = \frac{\prod_{k=1}^{D}\left(a_k w_k^{1-\nu}\right)^{1/\nu}}{\Phi^\circ(\infty; \boldsymbol{0}, \Sigma)\nu^{D-1}|\Sigma|^{1/2}\pi^{D/2}2^{(2-\nu)/2}} \left\{\left[(\boldsymbol{w}^\circ)^{1/\nu}\right]^\top \Sigma^{-1}\left[(\boldsymbol{w}^\circ)^{1/\nu}\right]\right\}^{-(\nu+D)/2} \Gamma\left(\tfrac{\nu+D}{2}\right), \quad \boldsymbol{w} = \tfrac{\boldsymbol{x}}{\|\boldsymbol{x}\|}.
$$

## D.3 Partial derivatives of the exponent function for the truncated extremal-t processes

**Lemma 4.** *The partial derivatives of the exponent function of the truncated extremal-t process are given by*

$$-V_{B_k}(\boldsymbol{x}) = \frac{\prod_{i \in B_k}\{a_i x_i^{1-\nu}\}^{1/\nu} 2^{(\nu-2)/2}\Gamma\left(\frac{k+\nu}{2}\right)\nu^{1-k}}{\Phi^\circ(\infty;0,\Sigma)\pi^{k/2}|\Sigma_{B_k B_k}|^{1/2} Q_{\Sigma_{B_k B_k}}\left(\boldsymbol{x}^\circ_{B_k}\right)^{k+\nu}}$$

$$\times T^\circ_{\nu+k}\left(\frac{\left(\boldsymbol{x}^\circ_{\overline{B}_k}\right)^{1/\nu}}{Q_{\Sigma_{B_k B_k}}\left(\boldsymbol{x}^\circ_{B_k}\right)}; \frac{\Sigma_{\overline{B}_k B_k}\Sigma^{-1}_{B_k B_k}\left(\boldsymbol{x}^\circ_{B_k}\right)^{1/\nu}}{Q_{\Sigma_{B_k B_k}}\left(\boldsymbol{x}^\circ_{B_k}\right)}, \frac{\Sigma_{\overline{B}_k \overline{B}_k} - \Sigma_{\overline{B}_k B_k}\Sigma^{-1}_{B_k B_k}\Sigma_{B_k \overline{B}_k}}{k+\nu}\right),$$

*where* $Q_\Sigma(\boldsymbol{x}) = \{\boldsymbol{x}^\top \Sigma^{-1}\boldsymbol{x}\}^{1/2}$.

*Proof.* Since the interest is in the partial derivatives of the exponent function $V(\boldsymbol{x})$ with respect to the first $k$ components, i.e., $\boldsymbol{x}_{B_k}$, the correlation matrix $\sigma$ is decomposed as

$$\Sigma = \begin{bmatrix} \Sigma_{B_k B_k} & \Sigma_{B_k \overline{B}_k} \\ \Sigma_{\overline{B}_k B_k} & \Sigma_{\overline{B}_k \overline{B}_k} \end{bmatrix}.$$

Plugging (S.7) into (12), we have

$$-V_{B_k}(\boldsymbol{x}) = \int_0^\infty r^k f_{W_{B_k}}(r\boldsymbol{x}_{B_k})\Pr(W_{\overline{B}_k} \in [0, r\boldsymbol{x}_{\overline{B}_k}] \mid W_{B_k} = r\boldsymbol{x}_{B_k})\mathrm{d}r$$

$$= \int_0^\infty r^k \frac{\phi\left(\left(r\boldsymbol{x}^\circ_{B_k}\right)^{1/\nu}; 0_k, \Sigma_{B_k B_k}\right)}{\Phi^\circ(\infty;0,\Sigma)} \prod_{i=1}^k \left\{\frac{a_i}{\nu}(rx_i^\circ)^{1/\Delta-1}\right\}$$

$$\times \Phi^\circ\left(\left(r\boldsymbol{x}^\circ_{\overline{B}_k}\right)^{1/\nu}; \Sigma_{\overline{B}_k}\Sigma^{-1}_{B_k B_k}\left(r\boldsymbol{x}^\circ_{B_k}\right)^{1/\nu}, \Sigma_{\overline{B}_k \overline{B}_k} - \Sigma_{\overline{B}_k B_k}\Sigma^{-1}_{B_k B_k}\Sigma_{B_k \overline{B}_k}\right)\mathrm{d}r$$

$$= \frac{\prod_{i \in B_k}\{x_i^{1/\nu-1}a_i^{1/\nu}\}}{\Phi^\circ(\infty;0,\Sigma)}\nu^{-k}\int_0^\infty r^{k/\nu}\phi\left(\left(r\boldsymbol{x}^\circ_{B_k}\right)^{1/\nu}; 0_k, \Sigma_{B_k B_k}\right)$$

$$\times \Phi^\circ\left(\left(r\boldsymbol{x}^\circ_{\overline{B}_k}\right)^{1/\nu}; \Sigma_{\overline{B}_k B_k}\Sigma^{-1}_{B_k B_k}\left(r\boldsymbol{x}^\circ_{B_k}\right)^{1/\nu}, \Sigma_{\overline{B}_k \overline{B}_k} - \Sigma_{\overline{B}_k B_k}\Sigma^{-1}_{B_k B_k}\Sigma_{B_k \overline{B}_k}\right)\mathrm{d}r$$

Using the notation $Q_\Sigma(\boldsymbol{x}) = \{\boldsymbol{x}^\top\Sigma^{-1}\boldsymbol{x}\}^{1/2}$ and applying the change of variable $t = r^{1/\nu}Q_{\Sigma_{B_k B_k}}(\boldsymbol{x}^\circ_{B_k})$, leads to

$$-V_{B_k}(\boldsymbol{x}) = \frac{\prod_{i \in B_k}\{x_i^{1/\nu-1}a_i^{1/\nu}\}}{\Phi^\circ(\infty;0,\Sigma)}\nu^{1-k}(2\pi)^{-k/2}|\Sigma_{B_kB_k}|^{-1/2}\int_0^\infty t^{k+\nu-1}Q_{\Sigma_{B_kB_k}}\left(\boldsymbol{x}^\circ_{B_k}\right)^{-k-\nu}e^{-t^2/2}$$

$$\times \Phi^\circ\left(\frac{\left(\boldsymbol{x}^\circ_{\overline{B}_k}\right)^{1/\nu}}{Q_{\Sigma_{B_kB_k}}\left(\boldsymbol{x}^\circ_{B_k}\right)}; \frac{\Sigma_{\overline{B}_kB_k}\Sigma^{-1}_{B_kB_k}\left(\boldsymbol{x}^\circ_{B_k}\right)^{1/\nu}t}{Q_{\Sigma_{B_kB_k}}\left(\boldsymbol{x}^\circ_{B_k}\right)}, \Sigma_{\overline{B}_k\overline{B}_k} - \Sigma_{\overline{B}_kB_k}\Sigma^{-1}_{B_kB_k}\Sigma_{\overline{B}_kB_k}\right)dt$$

$$= \frac{\prod_{i \in B_k}\{x_i^{1/\nu-1}a_i^{1/\nu}\}}{\Phi^\circ(\infty;0,\Sigma)}2^{(\nu-2)/2}\nu^{1-k}\pi^{-k/2}|\Sigma_{B_kB_k}|^{-1/2}Q_{\Sigma_{B_kB_k}}\left(\boldsymbol{x}^\circ_{B_k}\right)^{-k-\nu}\Gamma\left(\frac{k+\nu}{2}\right)$$

$$\times T^\circ_{\nu+k}\left(\frac{\left(\boldsymbol{x}^\circ_{\overline{B}_k}\right)^{1/\nu}}{Q_{\Sigma_{B_kB_k}}\left(\boldsymbol{x}^\circ_{B_k}\right)}; \frac{\Sigma_{\overline{B}_kB_k}\Sigma^{-1}_{B_kB_k}\left(\boldsymbol{x}^\circ_{B_k}\right)^{1/\nu}}{Q_{\Sigma_{B_kB_k}}\left(\boldsymbol{x}^\circ_{B_k}\right)}, \frac{\Sigma_{\overline{B}_k\overline{B}_k} - \Sigma_{\overline{B}_kB_k}\Sigma^{-1}_{B_kB_k}\Sigma_{\overline{B}_kB_k}}{k+\nu}\right),$$

which concludes the proof. $\square$

From Lemma 4, we can demonstrate that, for $k = 1, \ldots, D-1$ we have

$$\lim_{\boldsymbol{x}_{\overline{B}_k} \downarrow \boldsymbol{0}_{D_k}} V_{B_k}(\boldsymbol{x}) = \frac{\prod_{i \in B_k}\{a_i x_i^{1-\nu}\}^{1/\nu}2^{(\nu-2)/2}\Gamma\left(\frac{k+\nu}{2}\right)\nu^{1-k}}{\Phi^\circ(\infty;0,\Sigma)\pi^{k/2}|\Sigma_{B_kB_k}|^{1/2}Q_{\Sigma_{B_kB_k}}\left(\boldsymbol{x}^\circ_{B_k}\right)^{k+\nu}}$$

$$\times \lim_{\boldsymbol{x}_{\overline{B}_k} \downarrow \boldsymbol{0}_{D_k}} T^\circ_{\nu+k}\left(\frac{\left(\boldsymbol{x}^\circ_{\overline{B}_k}\right)^{1/\nu}}{Q_{\Sigma_{B_kB_k}}\left(\boldsymbol{x}^\circ_{B_k}\right)}; \frac{\Sigma_{\overline{B}_kB_k}\Sigma^{-1}_{B_kB_k}\left(\boldsymbol{x}^\circ_{B_k}\right)^{1/\nu}}{Q_{\Sigma_{B_kB_k}}\left(\boldsymbol{x}^\circ_{B_k}\right)}, \frac{\Sigma_{\overline{B}_k\overline{B}_k} - \Sigma_{\overline{B}_kB_k}\Sigma^{-1}_{B_kB_k}\Sigma_{\overline{B}_kB_k}}{k+\nu}\right)$$

$$= \frac{\prod_{i \in B_k}\{a_i x_i^{1-\nu}\}^{1/\nu}2^{(\nu-2)/2}\Gamma\left(\frac{k+\nu}{2}\right)\nu^{1-k}}{\Phi^\circ(\infty;0,\Sigma)\pi^{k/2}|\Sigma_{B_k,B_k}|^{1/2}Q_{\Sigma_{B_kB_k}}\left(\boldsymbol{x}^\circ_{B_k}\right)^{k+\nu}}$$

$$\times T^\circ_{\nu+k}\left(\boldsymbol{0}_{D_k}; \frac{\Sigma_{\overline{B}_kB_k}\Sigma^{-1}_{B_kB_k}\left(\boldsymbol{x}^\circ_{B_k}\right)^{1/\nu}}{Q_{\Sigma_{B_kB_k}}\left(\boldsymbol{x}^\circ_{B_k}\right)}, \frac{\Sigma_{\overline{B}_k\overline{B}_k} - \Sigma_{\overline{B}_kB_k}\Sigma^{-1}_{B_kB_k}\Sigma_{\overline{B}_kB_k}}{k+\nu}\right)$$

$$= \frac{\prod_{i \in B_k}\{a_i x_i^{1-\nu}\}^{1/\nu}2^{(\nu-2)/2}\Gamma\left(\frac{k+\nu}{2}\right)\nu^{1-k}}{\Phi^\circ(\infty;0,\Sigma)\pi^{k/2}|\Sigma_{B_k,B_k}|^{1/2}Q_{\Sigma_{B_kB_k}}\left(\boldsymbol{x}^\circ_{B_k}\right)^{k+\nu}} \times 0$$

$$= 0,$$

which implies that all densities on $\partial\Omega$ take the value zero.

### D.4 Exact simulation algorithm

Similarly to Section C.4, the exact simulation algorithm for truncated extremal-*t* process arises from Algorithm 2 in Dombry *et al.* (2016), which requires simulating a process $W/W(s_0)$ with distribution $P_{s_0}(A) = \mathrm{E}_{\boldsymbol{W}}[\mathbb{1}_{\{\boldsymbol{W}/W(s_0) \in A\}}W(s_0)]$ and where $\boldsymbol{W} = (W_0, W_1, \ldots, W_D) \equiv$ $(W(s_0), W(s_1), \ldots, W(s_D))$ is defined as in Theorem 3 to the exception of being $D+1$ dimensional. The distribution $P_{s_0}$ can be written as,

$$P_{s_0}(A) = \int_{\mathbb{R}_+^{D+1}} \mathbb{1}\{w_i/w_0 \in A_i, i = 1, \ldots, D\} w_0 dF_W(w)$$

$$= \int_{\mathbb{R}_+^{D+1}} \mathbb{1}\{w_i/w_0 \in A_i, i = 1, \ldots, D\} w_0 \frac{\phi\left((aw)^{1/\nu};0_{D+1},\Sigma\right)}{\Phi^\circ(\infty;0_{D+1},\Sigma)} \prod_{i=0}^{D} (a_i w_i)^{1/\nu-1} \frac{a_i}{\nu} dw,$$

where $a = (a_o, \ldots, a_D)$. Let $z = (z_0, z_1, \ldots, z_D)$, with $z_0 = 1$ and $z_i = w_i/w_0, i = 1, \ldots, D$. The above equation can be written as

$$P_{s_0}(A) = \int_{\mathbb{R}_+^{D+1}} \mathbb{1}\{z_i \in A_i, i = 1, \ldots, D\} w_0^{D+1} \frac{\phi\left((az)^{1/\nu};0_{D+1},\Sigma\right)}{\Phi^\circ(\infty;0_{D+1},\Sigma)} \prod_{i=0}^{D} (a_i w_0 z_i)^{1/\nu-1} \frac{a_i}{\nu} dw_0 dz_{-1}$$

$$= \int_{\mathbb{R}_+^{D+1}} \mathbb{1}\{z_i \in A_i, i = 1, \ldots, D\} w_0^{(D+1)/\nu} \frac{\phi\left((az)^{1/\nu};0_{D+1},\Sigma\right)}{\Phi^\circ(\infty;0_{D+1},\Sigma)} \prod_{i=0}^{D} (a_i z_i)^{1/\nu-1} \frac{a_i}{\nu} dw_0 dz_{-1}$$

Therefore, letting $\tilde{z} = (az)^{1/\nu}$, the density of the distribution $P_{s_0}(z)$ is proportional to

$$f(z) \propto \prod_{i=1}^{D} \left\{z_i^{1/\nu-1}\right\} \int_0^\infty w_0^{(D+1)/\nu} \exp\left\{-w_0^{2/\nu} \tilde{z}^\top \Sigma^{-1} \tilde{z}\right\} dw_0$$

$$= \prod_{i=1}^{D} \left\{z_i^{1/\nu-1}\right\} \int_0^\infty (2t)^{(D+1)/2} (\tilde{z}^\top \Sigma^{-1} \tilde{z})^{-(D+\nu+1)/2} e^{-t} \nu(2t)^{\nu/2-1} dt$$

$$\propto \prod_{i=1}^{D} \left\{z_i^{1/\nu-1}\right\} \left(\tilde{z}^\top \Sigma^{-1} \tilde{z}\right)^{-(D+\nu+1)/2} \int_0^\infty t^{(D+\nu-1)/2} e^{-t} dt$$

$$\propto \prod_{i=1}^{D} \left\{z_i^{1/\nu-1}\right\} \left(\tilde{z}^\top \Sigma^{-1} \tilde{z}\right)^{-(D+\nu+1)/2}$$

Recall that $B_D$ represents the set $\{1, \ldots, D\}$, given the block decomposition of the correlation matrix, $\Sigma = \begin{bmatrix} 1 & \Sigma_{0;B_D} \\ \Sigma_{0;B_D}^\top & \Sigma_{B_D;B_D} \end{bmatrix}$, the above equation can be written as

$$f(z) \propto \prod_{i=1}^{D} \left\{z_i^{1/\nu-1}\right\} \left[\frac{\tilde{z}_{B_D}/\tilde{z}_0 - \Sigma_{B_D;0}}{\Sigma_{B_D;B_D} - \Sigma_{B_D;0}\Sigma_{B_D;0}} \left(\tilde{z}_{B_D}/\tilde{z}_0 - \Sigma_{B_D;0}\right) + 1\right]^{-(D+\nu+1)/2},$$

which can be identified as the joint density function of $a_0 Z_i^\nu/a_i, i = 1, \ldots, D$, where $(Z_1, \ldots, Z_D)$ follows a multivariate truncated student-t distribution from below zero with scale parameter $(\Sigma_{B_D;B_D} - \Sigma_{B_D;0}\Sigma_{B_D;0}^\top)/(\nu + 1)$, location parameter $\Sigma_{B_D;0}$ and $\nu + 1$ degrees of freedom. Algorithm 2 gives the pseudo-code of the exact simulation of the truncated extremal-t processes at finite location $s_1, \ldots, s_D$.

---

**Algorithm 2:** Exact simulation of truncated extremal-t processes at locations $s = (s_1, \ldots, s_D)$.

---

**Input:** degree of freedom $\nu > 0$, $D \times D$ correlation matrix $\Sigma$
**Output:** $Z(s_j), j = 1, \ldots, D$.

1 Compute $a$ from (D.1).
2 Simulate $E_0 \sim \text{Exp}(1)$ and set $z = 1/E_0$.
3 Set $Y(s_1) = 1$ and simulate $Y(s)$ from a truncated-$t$ distribution with location parameter $\Sigma_{-1;1}$, scale matrix $(\Sigma - \Sigma_{-1;1}\Sigma_{-1;1}^\top)/(\nu + 1)$, and $\nu + 1$ degrees of freedom at location $s_i \neq s_1$.
4 Set $W(s_j) = a_0 Y^\nu(s_j)/a_j$ for $j = 1, \ldots, D$.
5 Set $Z(s_j) = W(s_j)z$ for $j = 1, \ldots, D$.
6 **for** $i = 2, \ldots, D$ **do**
7 $\quad$ Simulate $E_0 \sim \text{Exp}(1)$ and set $z = 1/E_0$.
8 $\quad$ **while** $z > Z(s_i)$ **do**
9 $\quad\quad$ Set $Y(s_j) = 1$ and simulate $Y(s_j), j \neq i$ from a truncated-$t$ distribution with location $\Sigma_{-j;j}$ and scale matrix $(\Sigma - \Sigma_{-j;j}\Sigma_{-j;j}^\top)/(\nu + 1)$, and $\nu + 1$ degrees of freedom.
10 $\quad\quad$ Let $W(s_j) = a_0 Y^\nu(s_j)/a_j, j = 1, \ldots, D$.
11 $\quad\quad$ **if** $W(s_j) \times z < Z(s_j), \forall j = 1, \ldots, i - 1$ **then**
12 $\quad\quad\quad$ Set $Z(s_j) = \max(Z(s_j), zW(s_j)), j = 1, \ldots, D$.
13 $\quad\quad$ Simulate $E_1 \sim \text{Exp}(1)$ and set $E_0 = E_0 + E_1$ and $z = 1/E_0$.

# E  Parameter estimation

## E.1  Proof of Theorem 4

The distribution of the r-Pareto process with risk functional $r$ is

$$\kappa_D(\cdot \cap \mathcal{A}_r)/\kappa_D(\mathcal{A}_r).$$

Suppose we have $\tilde{Y} \in \Omega$ follow the angular distribution $\mathcal{H}/D$ with support $\{x \in \Omega : \|x\|_1 = 1\}$, such that $1/D x^{-2} \mathrm{d}x \mathrm{d}\mathcal{H} = \mathrm{d}\kappa_D$, and $\tilde{R}$ follows a unit Pareto distribution, i.e., $\Pr[\tilde{R} > x] = 1/x, x > 1$, we have

$$\Pr[\tilde{R}\tilde{Y} \in \tilde{A}] = \frac{1}{D} \int_\Omega \int_1^\infty x^{-2} \mathbf{1}_{\{xw \in \tilde{A}\}} \mathrm{d}x \mathcal{H}(\mathrm{d}w).$$

If we have $\tilde{A} \subset \mathcal{A}_{r_0}$, then $xw \in \tilde{A}$ leads to $r_0(xw) = x\|w\|_1 = x > 1$. Therefore, we have

$$\Pr[\tilde{R}\tilde{Y} \in \tilde{A}] = \frac{1}{D} \int_\Omega \int_1^\infty x^{-2} \mathbf{1}_{\{xw \in \tilde{A}\}} \mathrm{d}x \mathcal{H}(\mathrm{d}w) = \frac{1}{D} \int_\Omega \int_0^\infty x^{-2} \mathbf{1}_{\{xw \in \tilde{A}\}} \mathrm{d}x \mathcal{H}(\mathrm{d}w) = \kappa_D(\tilde{A})/D.$$

As we have $\Pr(\tilde{R}\tilde{Y} \in \mathcal{A}_{r_0}) = \kappa_D(\mathcal{A}_{r_0})/D = 1/D$, it immediately follows that $\tilde{R}\tilde{Y}$ is distributed as the r-Pareto process with risk functional $r_0$. Let $\tilde{r}(\cdot) = c_0 r(\cdot)$, then $\tilde{r}$ will be also a risk functional that is convex and homogeneous of order 1. Thus, we have $\tilde{r}(e_i) = c_0/c_i \leq 1$. The complement set of $\mathcal{A}_{\tilde{r}}$ in $\mathbb{R}_+^D$, $\mathcal{A}_{\tilde{r}}^c = \{x \in \mathbb{R}_+^D : \tilde{r}(x) \leq 1\}$, is a convex set and contains points $e_i, i = 1, \ldots, D$ on its boundary as well as the origin $\mathbf{0}$ since any homogenous function of order 1 maps the origin to the origin. For $r_0$, we know that $\mathcal{A}_{r_0}^c$ is a convex set and contains $\mathbf{0}$ and $e_i, i = 1, \ldots D$ on its boundary, and $\mathcal{A}_{r_0}^c$ is the smallest convex set that contains points $e_i, i = 1, \ldots, D$ and $\mathbf{0}$. Therefore, we have $\mathcal{A}_{r_0}^c \subset \mathcal{A}_{\tilde{r}}^c$, which implies that $\mathcal{A}_{\tilde{r}} \subset \mathcal{A}_{r_0}$. If we have $\tilde{A} \subset \mathcal{A}_{\tilde{r}}$, then we also have $\tilde{A} \subset \mathcal{A}_{r_0}$. Moreover, we can get

$$\Pr[\tilde{Z}^{(r_0)} \in \tilde{A} | \tilde{Z}^{(r_0)} \in \mathcal{A}_{\tilde{r}}] = \frac{1/D \kappa_D(\tilde{A} \cap \mathcal{A}_{\tilde{r}})}{1/D \kappa_D(\mathcal{A}_{\tilde{r}})} = \frac{\kappa_D(\tilde{A} \cap \mathcal{A}_{\tilde{r}})}{\kappa_D(\mathcal{A}_{\tilde{r}})} \tag{S.8}$$

Since $\kappa_D$ is homogeneous of order -1 and $\mathcal{A}_{\tilde{r}} = 1/c_0 \mathcal{A}_r$, it follows that

$$\Pr[\tilde{Z}^{(r_0)} \in \tilde{A} | \tilde{Z}^{(r_0)} \in \mathcal{A}_{\tilde{r}}]$$

$$= \Pr[c_0 \tilde{Z}^{(r_0)} \in c_0 \tilde{A} | c_0 \tilde{Z}^{(r_0)} \in \mathcal{A}_r]$$

$$= \frac{\kappa_D(\tilde{A} \cap \mathcal{A}_r/c_0)}{\kappa_D(\mathcal{A}_r/c_0)} = \frac{c_0 \kappa_D(c_0 \tilde{A} \cap \mathcal{A}_r)}{c_0 \kappa_D(\mathcal{A}_r)} = \frac{\kappa_D(c_0 \tilde{A} \cap \mathcal{A}_r)}{\kappa_D(\mathcal{A}_r)}.$$

where $c_0 \tilde{A} \subset \mathcal{A}_r$ since $\tilde{A} \subset \mathcal{A}_{\tilde{r}}$. Therefore, we have $\tilde{Z}^{(r)} \overset{d}{=} c_0 \tilde{Z}^{(r_0)} \mid r\left(c_0 \tilde{Z}^{(r_0)}\right) > 1$.

## E.2 Comparing the spectral and pairwise composite likelihoods

Consider samples of size $n = 500$ from the Brown-Resnick model defined on a $15 \times 15$ grid (i.e., $D = 225$) with 6 parameter combinations taking $\lambda = 3, 6$ and $\vartheta = 0.5, 1, 1.5$. For the spectral likelihood (8), the threshold is set as $= 30 \times D$, resulting in about 16 exceedances while the weights $\zeta_{B_k}$ in the composite likelihood (6) take value 1 when the distance between two locations is less than $2\sqrt{2}$ and 0 otherwise, yielding 2,268 pairs in the likelihood evaluation. After replicating the estimation procedure 300 times, violin plots of the estimators $\widehat{\lambda}$ and $\widehat{\vartheta}$ are displayed in Figure 1, using the spectral (red) and pairwise composite likelihoods (blue), where the true parameter values are indicated by black dots. Results indicate that when in presence of relatively weak dependence (i.e., small $\lambda$ and large $\vartheta$), the bias exhibited by the spectral likelihood estimates seems to increase, overestimating the range parameter $\lambda$ and underestimating the smoothness parameter $\vartheta$. The presence of bias is not surprising since data with weak dependence possess more effective independent samples compared to cases with stronger dependence, and has been observed in previous studies (see e.g. Engelke *et al.*, 2015; Huser *et al.*, 2016). As expected, the pairwise likelihood estimator produces unbiased estimates but uses 500 observations against approximately 16 for the spectral likelihood. In terms of computation speed, the pairwise composite likelihood is about 476 times slower than the spectral likelihood in terms of core hours (average of 13.7 minutes with 16 CPU cores versus 9.2 seconds with 3 CPU cores). Overall, the spectral likelihood approach does produce estimates that are biased although those aren't very large, in particular for stronger dependence. As mentioned in Section 3.1 of the main manuscript, a censored likelihood approach could be taken at the expense of reducing the great computational gains. Furthermore, as dimension increases, the relative performance between the two approaches is expected to improve due to more available data (higher $u$, less bias).

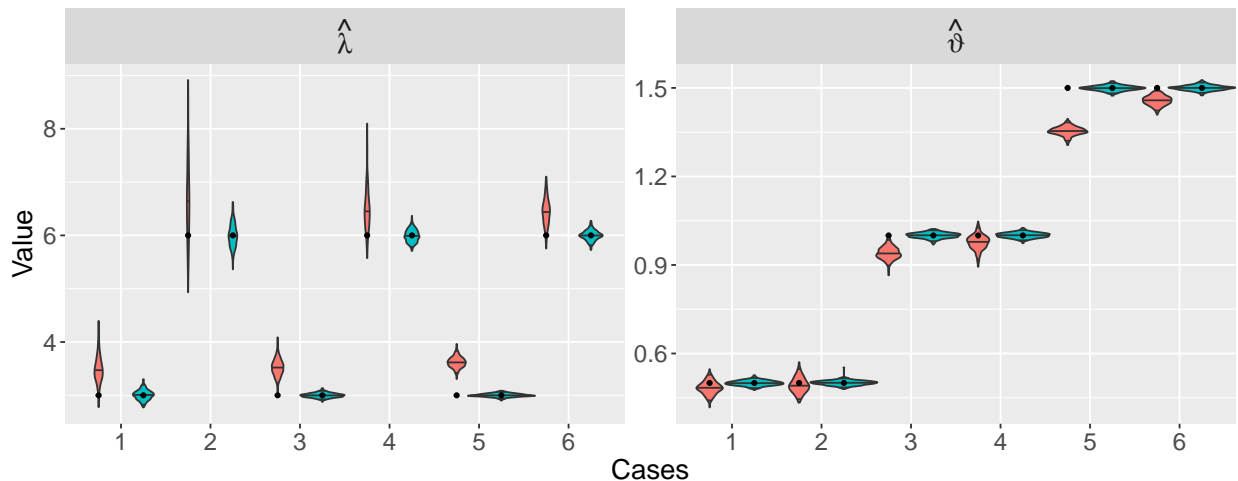

Figure 1: Violin plots for 300 replicate estimates of $\theta = (\lambda, \vartheta)$ for the Brown-Resnick process with $\lambda = 3, 6$ and $\vartheta = 0.5, 1, 1.5$ using the spectral (red) and pairwise (blue) likelihoods. Black dots indicate the true values.

### E.3 Comparing the spectral likelihood and score matching approach

We now generate $2,000$ samples from the truncated extremal-$t$ $r$-Pareto process on a $8 \times 8$ grid with 8 parameter combinations using $\lambda = (3, 5)$, $\vartheta = (1, 1.5)$, and $\nu = 2, 3$ and $L_1$ and $L_3$ norms as risk functionals. The model is fitted using both the score matching and spectral likelihood approaches with the outputs illustrated in Figure 4 when the $L_3$ and $L_1$-norm risk functional are used (respectively top and bottom two rows). We first note that there is barely any change in the performance of the estimators depending on the risk functional. No matter the norm used ($L_3$ or $L_1$), both estimation methods yield almost unbiased estimates for the smoothness parameter $\vartheta$. Some bias is however present for the estimation of the range parameter $\lambda$, with the spectral likelihood producing about 20% relative bias, while the score matching method again shows numerical instability. Again, the spectral likelihood shows greater computational efficiency, being about 10 times faster that the score matching method when using the $L_3$-norm risk functional (126 seconds on 3 CPU cores).

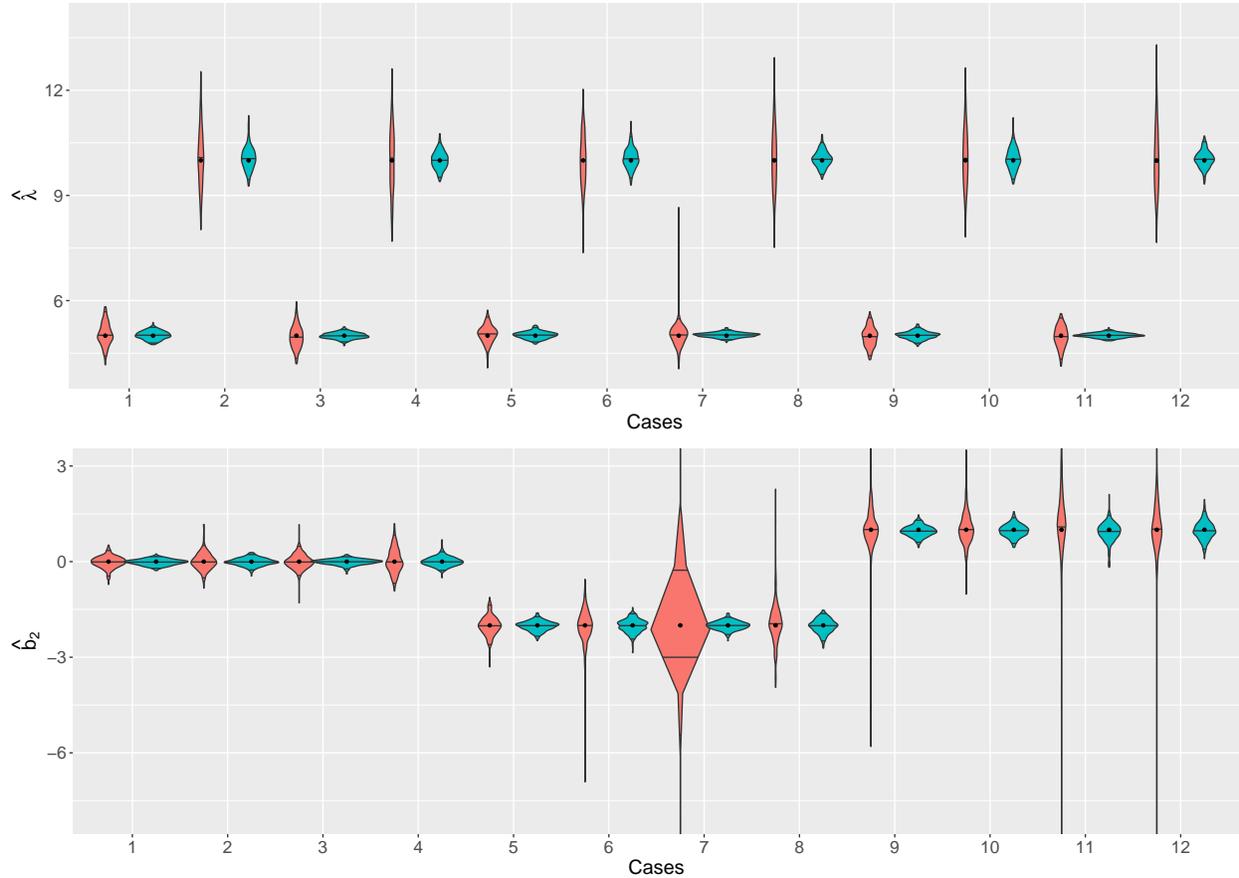

Figure 2: Violin plots for the score matching (red) and spectral likelihood (blue) estimates of $\lambda$ (top) and $b_2$ (bottom) for the skewed Brown-Resnick $r$-Pareto process with $L_3$ norm risk functional and the same parameter combinations as in Figure 4 of the main manuscript. Black dots indicate the parameter true values.

## F  Real-world applications

See Figure 5.

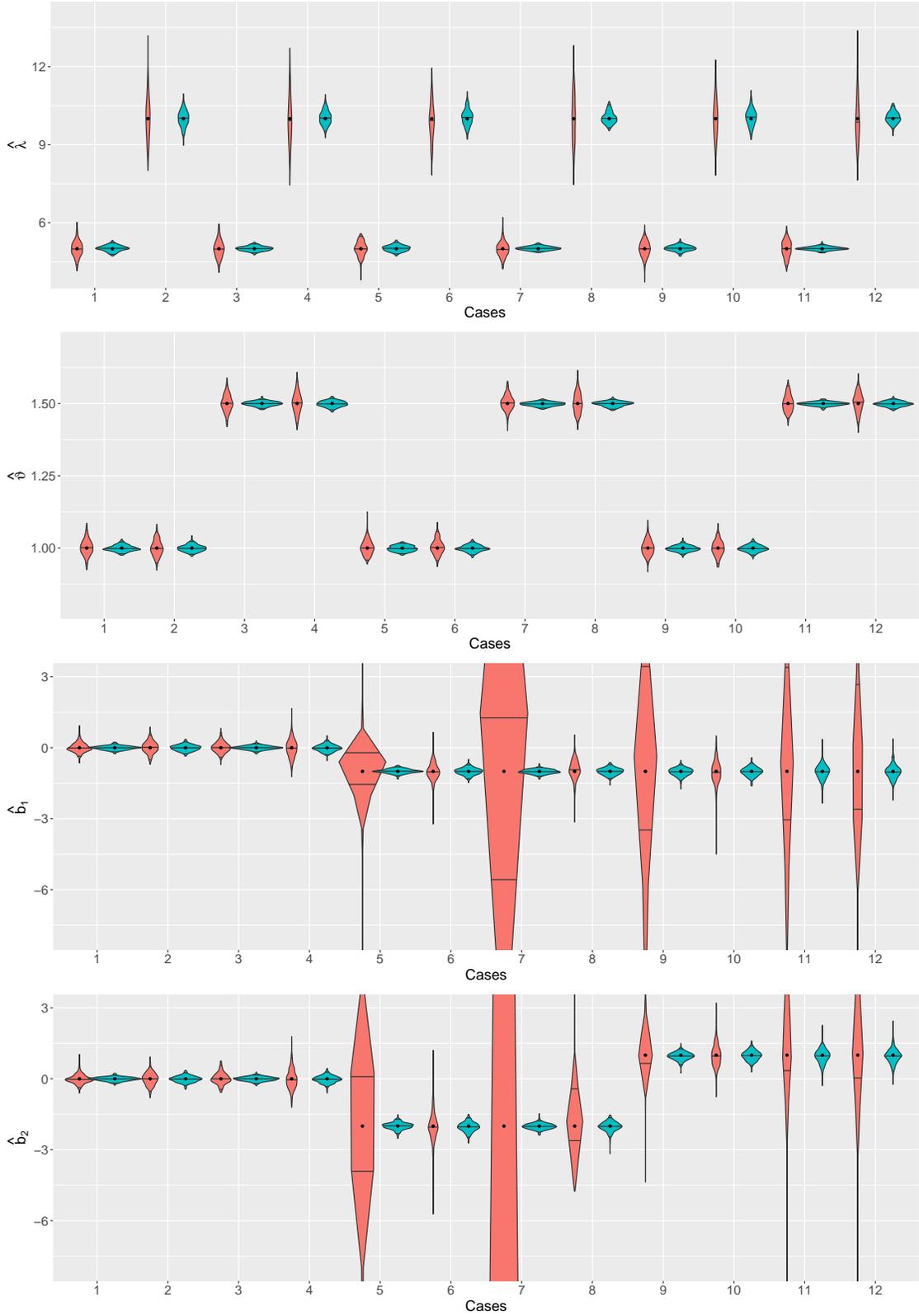

Figure 3: Violin plots for the score matching (red) and spectral likelihood (blue) estimates of $\boldsymbol{\theta} = (\lambda, \vartheta, b_1, b_2)$ (top to bottom) for the skewed Brown-Resnick $r$-Pareto process with $L_1$ norm risk functional and the same parameter combinations as in Figure 4 of the main manuscript. Black dots indicate the parameter true values.

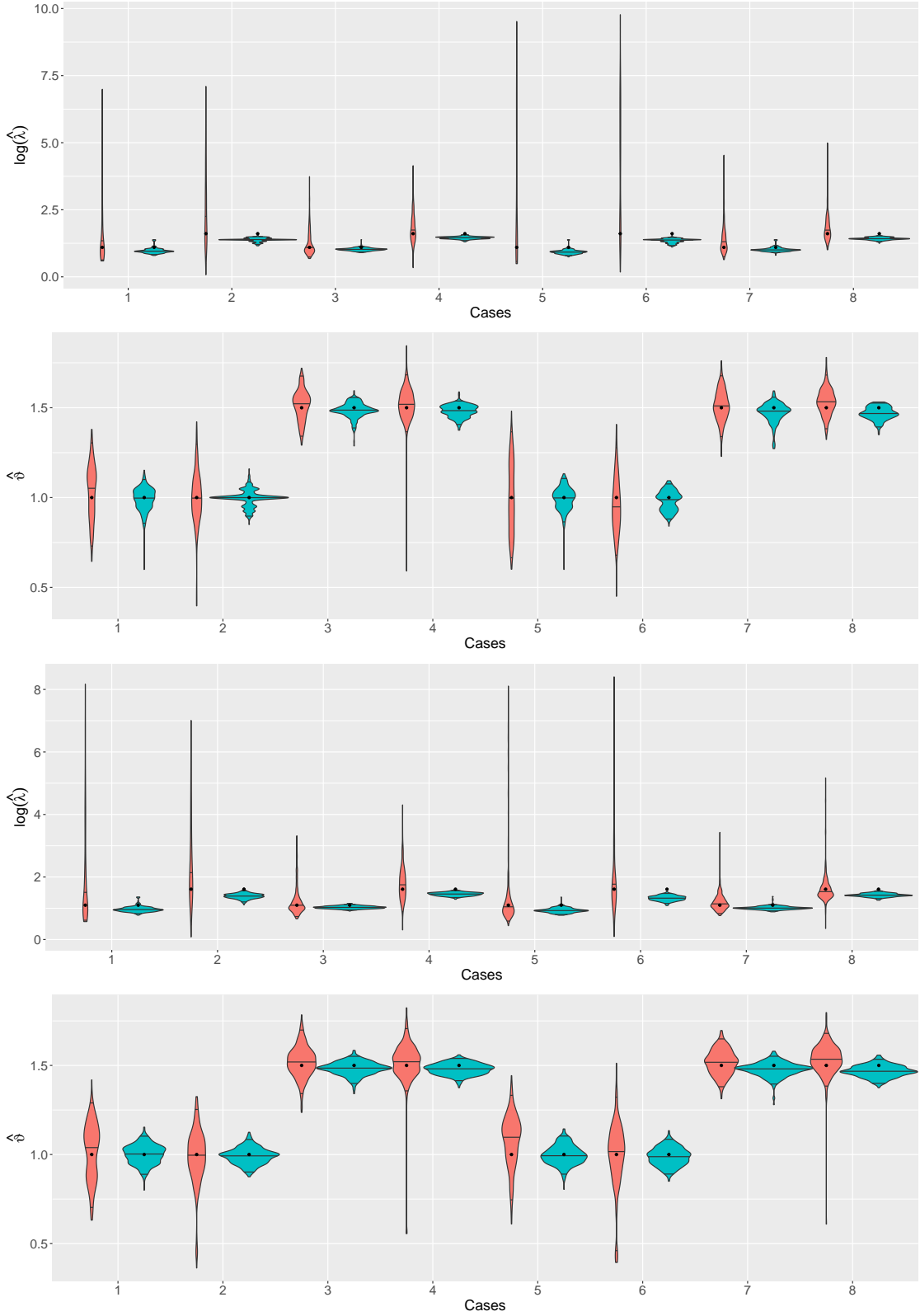

Figure 4: Violin plots for the score matching (red) and spectral likelihood (blue) estimates of $\boldsymbol{\theta} = (\log(\lambda), \vartheta)$ for the truncated extremal-$t$ $r$-Pareto processes with $L_3$ norm (top rows) and $L_1$ norm (bottom rows) risk functional, where $\lambda = c(3, 5), \vartheta = (1, 1.5)$ and $\nu$ is fixed to be either 2 (case 1–4) or 3 (case 5–8). Black dots indicate the true values.

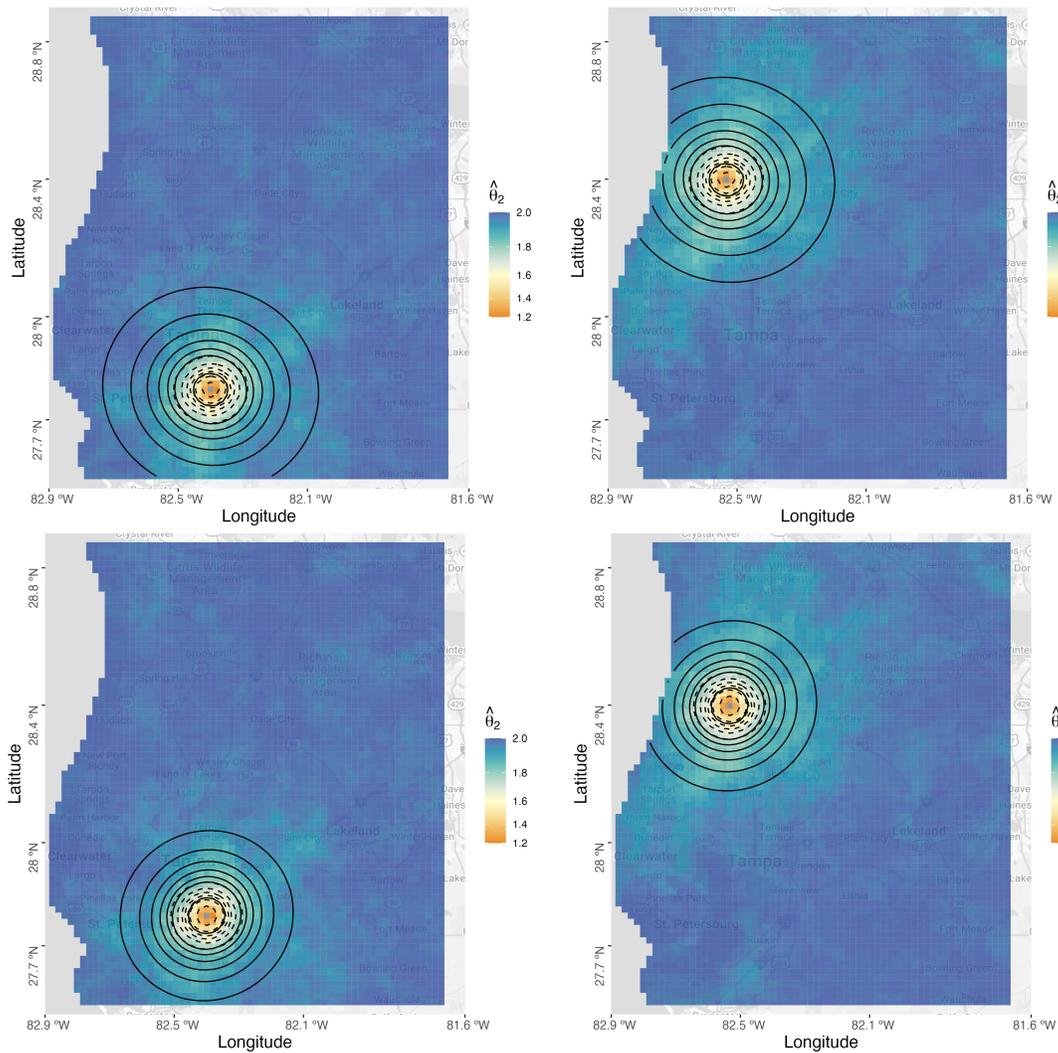

Figure 5: Maps of bivariate empirical extremal coefficients (shading) with respect to two different reference points (columns), and contours of the extremal coefficient of the Brown-Resnick $r$-Pareto model fitted using the spectral likelihood (solid line) and score matching approach (dashed line) on the Florida rainfall data. The top and bottom rows respectively consider the $L_1$ and $L_\infty$ norm risk functional.

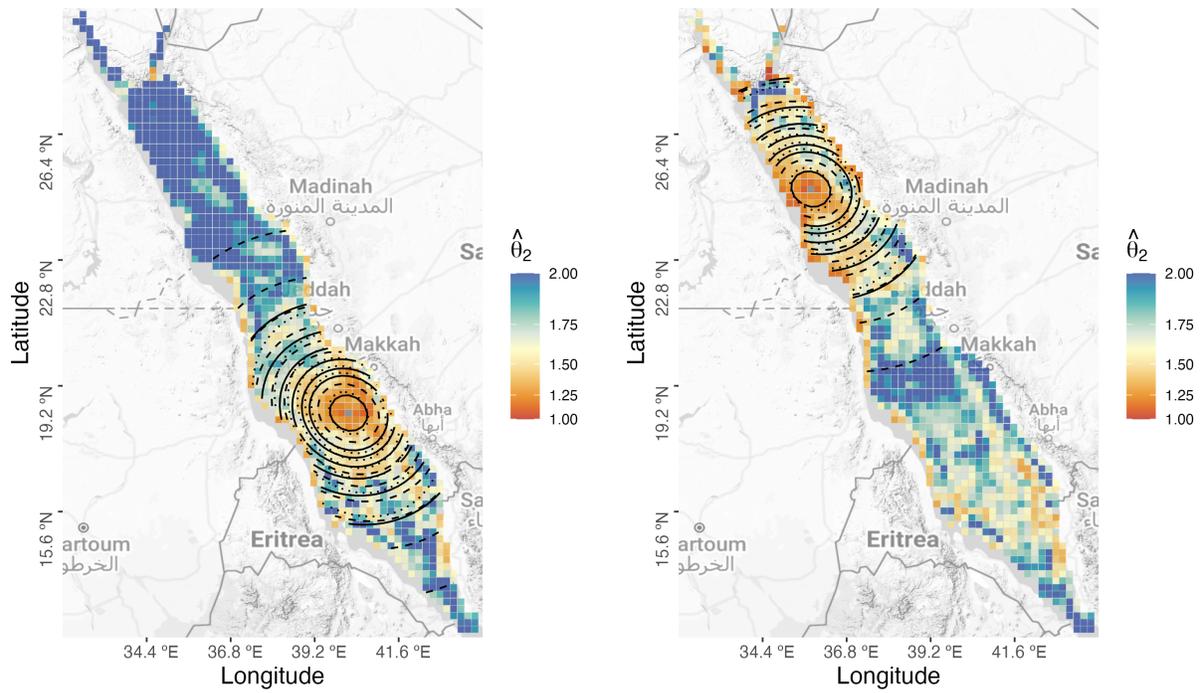

Figure 6: Maps of the bivariate empirical extremal coefficient (shading) with respect to two reference locations (left and right panels) with contours curves of the Brown-Resnick extremal coefficient fitted using the spectral likelihood (solid), pairwise composite likelihood (dotted) and Vecchia approximations (dashed).

26